\documentclass[aps,pra,twocolumn,showpacs,floatfix]{revtex4}
\usepackage{graphicx}
\usepackage{nicefrac}
\usepackage{amsmath}
\usepackage{amsfonts}
\usepackage{amssymb}
\usepackage{amsthm}
\usepackage{epsf}
\usepackage{bm}
\usepackage{bbm}
\usepackage{longtable}

\usepackage{dcolumn}

\newcolumntype{.}{D{x}{}{-1}}
\newcommand{\balpha}{\bm{\alpha}}
\newcommand{\bgamma}{\bm{\gamma}}
\newcommand{\bsigma}{\bm{\sigma}}
\newcommand{\bmu}{\bm{\mu}}
\newcommand{\bfx}{\bm{x}}
\newcommand{\bfp}{\bm{p}}
\newcommand{\bfq}{\bm{q}}
\newcommand{\bfr}{\bm{r}}
\newcommand{\bfc}{\bm{c}}
\newcommand{\hp}{\hat{\bm{p}}}

\newcommand{\lbr}{\langle}
\newcommand{\rbr}{\rangle}

\newcommand{\Za}{{Z\alpha}}

\newcommand{\vare}{\varepsilon}

\newcommand{\pr}{^{\prime}}
\newcommand{\ppr}{^{\prime\prime}}

\newcommand{\cross}[1]{#1\!\!\!/}

\newcommand{\SixJ}[6]{
        \left\{
        \begin{array}{ccc}
        #1  & #2  & #3 \\
        #4  & #5  & #6 \\
        \end{array}
        \right\}
        }

\begin{document}

\title{Self-energy correction to the hyperfine splitting\\
and the electron $\boldsymbol{g}$ factor in hydrogen-like ions}

\author{Vladimir A. Yerokhin}
\affiliation{Max--Planck--Institut f\"ur Kernphysik,
Postfach 10 39 80, 69029 Heidelberg, Germany}
\affiliation{Center
for Advanced Studies, St.~Petersburg State Polytechnical
University, Polytekhnicheskaya 29, St.~Petersburg 195251, Russia}

\author{Ulrich D. Jentschura}
\affiliation{Department of Physics,
Missouri University of Science and Technology,
Rolla, Missouri 65409-0640, USA}
\affiliation{Institut f\"ur Theoretische Physik,
Universit\"{a}t Heidelberg,
Philosophenweg 16, 69120 Heidelberg, Germany}

\begin{abstract}

The hyperfine structure (hfs) and the $g$ factor of a bound 
electron are caused by external magnetic fields.
For the hfs, the magnetic field is due to the nuclear spin.
A uniform-in-space and constant-in-time magnetic 
field is used to probe the bound-electron $g$ factor. 
The self-energy corrections to these effects are more difficult 
to evaluate than those to the Lamb shift. Here, we describe a numerical 
approach for both effects in the notoriously problematic 
regime of hydrogen-like bound systems with low nuclear charge numbers.
The calculation is nonperturbative in the binding Coulomb field.
Accurate numerical values for the remainder functions are
provided for $2P$ states and for $nS$ states with $n=1,2,3$.

\end{abstract}

\pacs{12.20.Ds, 31.30.Jv, 31.15.-p, 06.20.Jr}

\maketitle

%
%
\section{Introduction}

The interaction of a bound electron and an atomic nucleus is characterized by
the parameter $\Za$, where $Z$ is the nuclear charge number and $\alpha$ is
the fine-structure constant.  This universal ``coupling parameter'' sets the
scale for calculations of the radiative corrections to various bound-state
effects including the hyperfine structure (hfs) and the bound-electron $g$
factor. Traditionally, theoretical investigations 
of radiative corrections  in light systems
relied upon an expansion in powers of $Z\alpha$ and $\ln(Z\alpha)$.
However, today it is desirable to advance theory
beyond the predictive limits given by the highest available terms in the
$Z\alpha$-expansion. This can be done by carrying out calculations 
with using nonperturbative (in
$Z\alpha$) propagators. Such calculations demand rather sophisticated
numerical techniques, which were developed relatively recently.
Indeed, all-order calculations of the self-energy (SE) correction 
in the presence of a magnetic field  started in
the 1990s \cite{persson:96:hfs,shabaev:96:jetp,blundell:97:pra,%
blundell:97:prl,persson:97:g,sunnergren:98:pra}, extending during past years
to a wide range of reference states and nuclear charge numbers
\cite{beier:00:pra,yerokhin:01:hfs,sapirstein:01:hfs,yerokhin:02:prl,yerokhin:04,%
yerokhin:05:hfs,sapirstein:06:hfs,sapirstein:08:hfs}. 

Numerical calculations of the SE corrections are particularly difficult 
for low values of $Z$. This is mainly for two reasons. First, the goal of
the calculations is the contribution beyond the known $\Za$-expansion terms.
For the hfs, the higher-order effects are 
suppressed with respect to the leading correction
by a factor of $(\Za)^3$. For the $g$ factor, they enter only
at order of $(\Za)^5$ and thus become very
small numerically in the low-$Z$ region. Second, in actual calculations there
are additional cancellations arising at intermediate stages of
the numerical procedure. These cancellations become more severe
for smaller values of $Z$ and lead to further losses of accuracy. 

In this article, we treat the two most important example cases
of the bound-electon SE corrections in external magnetic fields:
the SE correction to the hfs
and the SE correction to the bound-electron $g$ factor.
We evaluate both of these corrections for the ground and
excited states of hydrogen and of light hydrogen-like ions. 
The first attempt at an all-order evaluation of the SE correction to
the hfs of hydrogen was made in Ref.~\cite{blundell:97:prl}.
Because of insufficient numerical accuracy,
the goal was reached in an indirect way: the known terms of the $\Za$
expansion were subtracted from numerically 
determined all-order results for $Z \ge 5$, and
the higher-order remainder was extrapolated down toward the 
desired value $Z=1$.
The accuracy of the numerical evaluation of the SE correction to the hfs was
improved by several orders of magnitude during the past years
\cite{yerokhin:01:hfs,yerokhin:05:hfs}. However, the precision obtained was
still insufficient for a direct determination of the higher-order SE remainder
at $Z=1$, and an extrapolation procedure had to be employed again. 

The studies \cite{yerokhin:01:hfs,yerokhin:05:hfs} reported results for the
higher-order contribution for the normalized difference of the $1S$ and $2S$ hfs
intervals in $^3{\rm He}^+$ and demonstrated a $2\sigma$ deviation of the
theoretical prediction from the experimental result
\cite{schluesser:69,prior:77}. The accuracy of the extrapolation procedure of
Refs.~\cite{yerokhin:01:hfs,yerokhin:05:hfs}, however, has recently become a subject of
some concern. In particular, an opinion was expressed
in Ref.~\cite{karshenboim:05:cjp} that the uncertainty of the extrapolation 
procedure should have been estimated as four times
larger than given in Refs.~\cite{yerokhin:01:hfs,yerokhin:05:hfs}, which would
have brought theory and experiment into agreement.

In our recent investigation \cite{yerokhin:08:prl}, we performed the first
direct, high-precision theoretical determination of the higher-order remainder
of the SE correction to the hfs of $1S$ and $2S$ states of hydrogen and
light hydrogen-like ions. Good agreement was observed with the previous extrapolated
values \cite{yerokhin:01:hfs,yerokhin:05:hfs},
but the accuracy was increased by several orders of magnitude. In the
present paper, we report the details of this calculation and extend it
to the higher excited states ($3S$, $2P_{1/2}$, and
$2P_{3/2}$). 

The SE correction to the bound-electron $g$ factor is of particular importance
because it is used in the determination of the electron mass value from the
experimental results for the $g$ factor of light hydrogen-like ions
\cite{mohr:08:rmp}. Already at the present level of experimental accuracy, 
calculations of the bound-electron $g$ factor should be performed
to all orders in $\Za$. A number of all-order evaluations of the SE
correction to the $g$ factor have been accomplished during last years 
\cite{persson:97:g,blundell:97:pra,beier:00:pra,yerokhin:02:prl,yerokhin:04},
which resulted in an improvement of the precision of the electron mass value. 
However, in order to match the $10^{-12}$ level of accuracy
anticipated in future experiments on the helium ion \cite{quint:08}, 
the precision of numerical calculations of the SE correction should be
enhanced by several orders of magnitude. 

First results of our evaluation of the SE correction to the bound-electron $g$
factor for the $1S$ state of light hydrogen-like ions were reported in
Ref.~\cite{yerokhin:08:prl}. In the present investigation we extend our
calculation to the higher excited states ($2S$, $3S$, $2P_{1/2}$, and
$2P_{3/2}$) and to a wider region of the nuclear charge number $Z$. 
Relativistic units ($\hbar = c = m = 1$) and Heaviside charge units
$(\alpha = e^2/4\pi, e<0)$ are used throughout the paper.

Our investigations are organized as follows. In Sec.~\ref{general},
we discuss general formulas pertaining to the formulation of the 
effect within the formalism of quantum electrodynamics.
We continue with a detailed description of the numerical approach 
in Sec.~\ref{detailed}. Numerical results are presented in 
Sec.~\ref{numerical}. We conclude with a summary in Sec.~\ref{conclusions}.

%
%
\begin{figure}[thb]
\centerline{\includegraphics[width=0.8\linewidth]{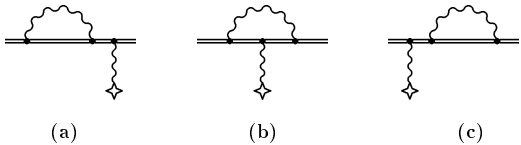}}
\caption{
Feynman diagrams representing the SE correction in the
presence of an external perturbing field. The double line indicates the bound electron
propagator, which is nonperturbative in the 
coupling parameter $Z\alpha$ and entails an arbitrary number of 
Coulomb interactions with the atomic nucleus.
The wavy line that ends with a cross denotes the interaction
with the perturbing potential $\delta V$. The latter is given by the 
magnetic field of the nucleus in the case of the hfs and 
by a constant external magnetic field in the case of the 
bound-electron $g$ factor. \label{fig:se} }
\end{figure}

%
%
\section{General formulas}
\label{general}

The SE correction in the presence of a binding Coulomb field and an
additional perturbing potential $\delta V$ is graphically represented by
the Feynman
diagrams shown in Fig.~\ref{fig:se}. The general expression for them can be
conveniently split into three parts \cite{shabaev:02:rep},
\begin{equation}\label{eq1}
    \Delta E_{\rm SE} = \Delta E_{\rm ir}+\Delta E_{\rm red}+\Delta E_{\rm
    ver}\,,
\end{equation}
which are referred to as the {\em irreducible}, the {\em reducible}, and 
the {\em vertex} contribution, respectively. 

The vertex contribution is induced by the diagram in Fig.~\ref{fig:se}(b). It
can be expressed as
\begin{align} \label{eq2}
&\Delta E_{\rm ver} = \ \frac{i}{2\pi}\int_{-\infty}^{\infty} d\omega\,
  \nonumber \\
& \times   \sum_{n_1n_2}
    \frac{\lbr n_1|\delta V|n_2\rbr\,
            \lbr an_2| I(\omega)|n_1a\rbr}
                   {[\vare_a-\omega-\vare_{n_1}(1-i 0)]
                   [\vare_a-\omega-\vare_{n_2}(1-i 0)]}\,.
\end{align}
Here, $I$ is the operator of the electron-electron interaction
\begin{equation}\label{eq3}
 I(\omega) = e^2\,\alpha_{\mu}\alpha_{\nu}\,D^{\mu\nu}(\omega)\,,
\end{equation}
where $D^{\mu\nu}$ is the photon propagator and ${\alpha}^{\mu} = (1,
\balpha)$ are the Dirac matrices.
The sums over $n_1$ and $n_2$ involve both the positive-energy 
discrete and continuous spectra and the 
negative-energy continuous spectrum.

The irreducible contribution is induced by a part of the diagrams in
Fig.~\ref{fig:se}(a) and (c) that can be expressed in terms of the
first-order perturbation of the reference-state wave function by $\delta V$,
\begin{equation}\label{eq4}
|\delta a\rbr = 
\sum_{\scriptsize \begin{array}{cc} n \\[-0.5ex]
{\vare_n \ne \vare_a} \end{array}}
\frac{|n\rbr \lbr n|\delta V|a\rbr}{\vare_a-\vare_n} \,.
\end{equation}
The expression for the irreducible contribution is
\begin{equation}\label{eq5}
\Delta E_{\rm ir} = \lbr \delta a|\gamma^0
    \widetilde{\Sigma}(\vare_a) |a\rbr
              + \lbr a|\gamma^0 \widetilde{\Sigma}(\vare_a) |\delta a\rbr\,,
\end{equation}
where $\widetilde{\Sigma} = \Sigma -\delta m$, $\delta m$ is the
one-loop mass counterterm, and $\Sigma$ is the one-loop SE
operator,
\begin{eqnarray} \label{eq6}
 \Sigma(\vare,\bfx_1,\bfx_2) &=&
2\,i\alpha\,\gamma^0 \int_{-\infty}^{\infty} d\omega\,
      \alpha_{\mu}\,
 \nonumber \\ && \times
         G(\vare-\omega,\bfx_1,\bfx_2)\, \alpha_{\nu}\,
    D^{\mu\nu}(\omega,\bfx_{12})\,
         \,.
\end{eqnarray}
In the above, $G$ denotes the Dirac Coulomb Green function 
$G(\vare) = [\vare-{\cal H}(1-i 0)]^{-1}$, $\cal H$ is 
the Dirac Coulomb Hamiltonian, and $\bfx_{12} = \bfx_1-\bfx_2$. 

The reducible contribution  is induced by a part of diagrams in
Fig.~\ref{fig:se}(a) and (c) that can be expressed in terms of the
first-order perturbation of the reference-state energy.
It reads
\begin{equation}\label{eq8}
    \Delta E_{\rm red} = \delta \vare_a\, \lbr a| \gamma^0 \left.
    \frac{\partial}{\partial \vare} \Sigma(\vare) \right|_{\vare = \vare_a} |
            a\rbr\,,
\end{equation}
where $\delta \vare_a = \lbr a|\delta V|a\rbr$. 

Up to now we did not specify the particular form of the perturbing potential
$\delta V$, assuming only its locality. In the following, we consider two
particular choices of $\delta V$, both representing interactions with the
magnetic field: the hfs interaction and the interaction with the magnetic
field (the Zeeman effect). In the case of the hfs interaction, the perturbing
potential has the form of the Fermi-Breit interaction 
\begin{equation}
\label{eq9}
    V_{\rm FB}(\bfr) =
      \frac{|e|}{4\pi}\,\frac{\balpha\cdot[\bmu\times\bfr]}{r^3}\,,
\end{equation}
(where $\bmu$ denotes the nuclear magnetic moment)
and the reference-state wave function $|a\rbr$ is the wave
function of the coupled system (electron$+$nucleus),
\begin{equation}\label{eq10}
    |a \rbr \to  |FM_FIj\rbr =  \sum_{M_Im_a} C^{FM_F}_{IM_I j_a
        m_a}\, |IM_I\rbr \,|j_a m_a\rbr\,,
\end{equation}
where $|IM_I\rbr$ denotes the nuclear wave function, $|j_a
m_a\rbr$ is the electron wave function, $F$ is the total momentum
of the atom, and $M_F$ is its projection. The nuclear variables can be
separated out by using the standard technique of the Racah algebra. 
It can be demonstrated \cite{yerokhin:05:hfs} that the general
formulas (\ref{eq1})--(\ref{eq8}) yield contributions to the hfs
if one employs an electronic perturbing potential of the form
\begin{equation}\label{eq11}
   \delta V_{\rm hfs}(\bfr) =
       \frac{E_F}{C_{\rm hfs}}\, \frac{[\bfr \times
            \balpha]_z}{r^3}\,,
\end{equation}
and takes the reference-state wave function to be the electronic wave function with
the momentum projection $m_a = \tfrac12$,
\begin{equation}\label{eq12}
    |a\rbr = |j_a\,\tfrac12\rbr\,.
\end{equation}
In the above, $E_F$ denotes the nonrelativistic limit of the expectation 
value of the Fermi-Breit operator on the reference state
and the prefactor $C_{\rm hfs}$ is given by
\begin{equation}\label{eq13}
C_{\rm hfs} = m^2\, (\Za)^3 \, 
\frac{{\rm sign}(\kappa_a)}{n_a^3 \, (2\kappa_a+1) \, (\kappa_a^2-\tfrac14)}\,,
\end{equation}
where $\kappa_a$ is the Dirac quantum number of the reference state and $n_a$ is
its principal quantum number. 

In the case of the Zeeman splitting, the perturbing potential is 
\begin{equation}
V_{\rm Zee}(\bfr) = -e\,\balpha \cdot {\bm A}(\bfr)\,,
\end{equation}
where  ${\bm A}(\bfr) = \tfrac12 \, [{\bm B}\times \bfr]$
is the vector potential. In practical calculations, corrections to the Zeeman
splitting are convenienly expressed in terms of corrections to the $g$
factor. It can be easily shown (see Ref.~\cite{yerokhin:04} for details)
that the general formulas
(\ref{eq1})-(\ref{eq8}) yield contributions to the electronic $g$ factor if
one employs a perturbing potential of the form
\begin{equation}\label{eq14}
   \delta V_{g}(\bfr) =
       2 m\, [\bfr \times \balpha]_z\,,
\end{equation}
and the reference-state wave function with the momentum projection $m_a=1/2$. 
The $g$-factor perturbing potential (\ref{eq14})
differs from the hfs potential (\ref{eq11}) only by the power of $r$ and
the prefactor.

In the following, all explicit formulas for individual contributions
will be presented for the case
of the hfs. When working in the coordinate representation, the corresponding
formulas for the  $g$ factor can be obtained by an obvious substitution. In
momentum space, the formulas for the hfs and for the $g$ factor are different. 
Our present approach to the evaluation of the SE correction to the $g$ factor
closely follows the one of Ref.~\cite{yerokhin:04} and is therefore not 
described separately.

%
%
\section{Detailed analysis}
\label{detailed}

%
%
\subsection{Orientation}

The general formulas presented in the
previous section for individual contributions 
are both ultraviolet (UV) and infrared (IR) divergent. In order to
obtain expressions suitable for numerical evaluation, a careful
re-arrangement of contributions is needed, together with a covariant
regularization of divergences. The calculation of the irreducible contribution
(\ref{eq5}) can be reduced to an evaluation of a non-diagonal matrix element of the
first-order SE operator (\ref{eq6}). Its renormalization is well
known and does not need to be discussed here. The numerical evaluation of the
irreducible contribution was performed by a generalization of the approach of
Refs.~\cite{jentschura:99:prl,jentschura:01:pra}, with the use of a
closed-form analytic representation of the perturbed wave function $|\delta
a\rbr$ obtained in Ref.~\cite{shabaev:91:jpb} (see also Ref.~\cite{shabaev:03:PSAS}).  

The evaluation of the reducible and the vertex contribution is carried out
after splitting them into several parts,
\begin{eqnarray}
\Delta E_{\rm red} &=&  \Delta E^{(a)}_{\rm red} + \Delta E^{(0)}_{\rm red}
   + \Delta E^{(1+)}_{\rm red} \,,
   \\
\Delta E_{\rm ver} &=& \Delta E^{(a)}_{\rm ver} + \Delta E^{(0)}_{\rm ver}
   + \Delta E^{(1)}_{\rm ver}+ \Delta E^{(2+)}_{\rm ver} \,,
\end{eqnarray}
where the upper index $(a)$ labels the contributions induced by the
reference-state part of the electron propagators and the other indices specify the
total number of interactions with the binding field in the electron propagators
[the index $(i+)$ labels the terms generated by $\ge$$i$ such interactions].

%
%
\subsection{Reference-state contribution}
\label{refstatecontrib}

The reference-state contributions $\Delta E^{(a)}_{\rm red}$ and  $\Delta
E^{(a)}_{\rm ver}$ are separately IR divergent. The divergences disappear
when the contributions are regularized in the same way and evaluated together.
Let us now demonstrate the cancellation of the IR divergences and obtain
the finite residual. The part of the vertex and reducible contributions induced
by the intermediate states degenerate in energy with the reference state  is
\begin{align} \label{d2}
\Delta E^{(a)} &\ \equiv \Delta E^{(a)}_{\rm ver}+ \Delta E^{(a)}_{\rm red} = 
  \frac{i}{2\pi}\int_{-\infty}^{\infty} d\omega \frac1{(\omega-i0)^2}\,
\nonumber \\ & \times
         \Bigl[  \sum_{\mu_{a\pr}\mu_{a\ppr} }
                \lbr a\pr |\delta V | a\ppr \rbr
                \lbr a a\ppr|I(\omega)| a\pr a\rbr
\nonumber \\ & 
- \sum_{\mu_{a\pr}}\lbr a |\delta V | a \rbr
                \lbr a a\pr|I(\omega)| a\pr a\rbr
                \Bigl]\,,
\end{align}
where $a$ is the ``true'' reference state and $a\pr$ and $a\ppr$ label 
the intermediate states that are degenerate with the reference state in energy 
and have momentum projections $\mu_{a\pr}$ and $\mu_{a\ppr}$, respectively. 
(The intermediate states degenerate with the reference state 
in energy but of opposite parity do not induce any IR divergences because 
of the orthogonality of the wave functions. However, in practical
calculations we find it convenient to treat all the degenerate states 
on the same  footing.)
For simplicity, we now consider the photon propagator in the Feynman gauge. Then,
the operator of the electron-electron interaction $I$ takes the form
\begin{equation}
 I(\omega) = \alpha\,\alpha_{\mu}\alpha^{\mu}\,D(\omega,x_{12})\,,
\end{equation}
where 
\begin{equation}\label{d4}
D(\omega,x_{12})  = -4\pi\int \frac{d \bm k}{(2\pi)^3} \frac{\exp(i\bm{k} \cdot
            \bfx_{12})}{\omega^2-\bm{k}^2-\mu^2+i0} \,,
\end{equation}
with $\mu$ being the photon mass, which regularizes the IR divergences. It can be
seen that all divergences in  Eq.~(\ref{d2}) originate from an integral of
the form
\begin{equation}
J = \frac{i}{2\pi}\int_{-\infty}^{\infty} 
d\omega\, \frac{D(\omega,x_{12})}{(\omega-i0)^2}\,,
\end{equation}
We now
substute Eq.~(\ref{d4}) into the above expression, twice perform an
integration by parts, evaluate the $\omega$ integral by Cauchy's
theorem, and obtain
\begin{equation}
J = -\frac{1}{\pi} \int_0^{\infty}dk\, \frac{\cos kx_{12}}{\sqrt{k^2+\mu^2}}\,.
\end{equation}
Adding and subtracting $\cos k$ in the numerator of the integrand, we separate
the above expression into two parts,
the first of which is convergent when $\mu\to 0$
while the other (divergent part) does not depend on $x_{12}$. 
Setting $\mu=0$ in the convergent
part and evaluating the integral, we obtain
\begin{equation}
J = \frac{1}{\pi} \ln x_{12}- 
\frac1{\pi}\int_0^{\infty}dk\, \frac{\cos k}{\sqrt{k^2+\mu^2}}\,.
\end{equation}
The divergent part of $J$ does not depend on the radial variables and, being
substituted into Eq.~(\ref{d2}), leads to a vanishing contribution. We thus
obtain
\begin{align} \label{d6}
\Delta E^{(a)} &\ = 
  \frac{\alpha}{\pi}
         \Bigl[  \sum_{\mu_{a\pr}\mu_{a\ppr} }
                \lbr a\pr |\delta V | a\ppr \rbr
                \lbr a a\ppr|\alpha_{\mu}\alpha^{\mu}\,\ln x_{12}| a\pr a\rbr
\nonumber \\ & 
- 
\sum_{\mu_{a\pr}}\lbr a |\delta V | a \rbr
                \lbr a a\pr|\alpha_{\mu}\alpha^{\mu}\,\ln x_{12}| a\pr a\rbr
                \Bigl]\,.
\end{align}
We note that in the case when the perturbing potential $\delta V$ is spherically
symmetric, the reference-state contribution $\Delta E^{(a)}$ vanishes as
$\lbr a\pr |\delta V | a\ppr \rbr = \delta_{\mu_{a\ppr}\mu_{a\pr}}
\lbr a |\delta V | a \rbr$. In our
case, however, $\delta V$ represents an interaction with the magnetic field, so
that $\Delta E^{(a)}$ induces a finite contribution. 

In our practical calculations, the reference-state contribution was separated from
the vertex and reducible parts by introducing point-by-point subtractions 
from the electron propagators in
the integrands and was calculated separately according to Eq.~(\ref{d6}).

%
%
\subsection{Zero-potential parts}
\label{zeropotredver}

The zero-potential parts $\Delta E^{(0)}_{\rm red}$ and $\Delta E^{(0)}_{\rm ver}$ are
separately UV divergent. They are covariantly regularized by working in
an extended number of dimensions ($D=4 - 2 \epsilon$)
and calculated in momentum space. The
elimination of UV divergences in the sum of the reducible and the
vertex contributions is well documented in the literature (see, e.g., 
Ref.~\cite{yerokhin:99:sescr}), so here we operate with the 
{\it renormalized} SE and vertex operators, assuming that all UV
divergences are already cancelled out. 

The zero-potential contribution to the reducible part is simple. It is given by
\begin{align}\label{e1}
    \Delta E_{\rm red}^{(0)} =&\ \lbr a|\delta V|a\rbr \, 
\nonumber \\ \times &
 \int \frac{d\bfp}{(2\pi)^3}\, \overline{\psi}_a(\bfp)\,
   \left.
    \frac{\partial}{\partial p^0}\, \Sigma_R^{(0)}(p) \right|_{p^0 = \vare_a} 
            \psi_a(\bfp)\,,
\end{align}
where $\overline{\psi} = \psi^{\dag} \gamma^0$ is the Dirac adjoint.
The derivative of the renormalized free SE operator $\Sigma_R^{(0)}$
can be expressed as a linear combination of 3 matrix structures, 
$\cross{p} \equiv \gamma^\mu p_\mu$,
$\gamma^0$, and the unity matrix $I$,
\begin{align} 
\left.
\frac{\partial\Sigma^{(0)}_R(p)}{\partial p^0} 
 \right|_{p^0 = \vare_a} 
  = -\frac{\alpha}{4\pi}
      \left[\frac{\cross{p}}{m^2}\, a_1(\rho)
      + \gamma_0\, a_2(\rho)+ I\,a_3(\rho) \right] ,
\end{align}
where $\rho = (m^2-p^2)/m^2 = (m^2-\vare_a^2+\bfp^2)/m^2$ and $a_i(\rho)$ are
scalar functions, whose explicit expression is given by Eqs.~(53)-(55) of 
Ref.~\cite{yerokhin:99:sescr}. Integrating over angular variables, we
immediately have
\begin{align} 
\Delta E_{\rm red}^{(0)}&\  =
\lbr a|\delta V|a\rbr \, 
\left( - \frac{\alpha}{4\pi}\right)\, 
\int\limits_0^{\infty} \frac{p_r^2\, dp_r}{(2\pi)^3}\,
         \nonumber \\ & \times
        \Bigl\{ a_1(\rho) [\vare_a(g_a^2+f_a^2)+2p_r g_af_a ]
         \nonumber \\
&  + a_2(\rho) ( g_a^2+f_a^2)
    + a_3(\rho) (
        g_a^2-f_a^2) \Bigr\} \,,
\end{align}
where $p_r = |\bfp|$ and  $g_a = g_a(p_r)$ and $f_a = f_a(p_r)$ are the upper
and the lower components of the reference-state wave function in the momentum
space. 

The zero-potential vertex part of the SE hfs correction 
is induced by the hfs potential $\delta V_{\rm hfs}$ 
inserted in the free SE loop. The hfs potential~(\ref{eq11}) in the
momentum space takes the form
\begin{equation}
\label{VhfsMomentum}
\delta V_{\rm hfs}(\bfq) = 
 \frac{E_F}{C_{\rm hfs}}\, (-4\pi i)\,\frac{[\bfq\times\balpha]_z}{\bfq^2}\,.
\end{equation}
The zero-potential vertex part is then given by
\begin{align} \label{e6}
\Delta E^{(0)}_{\rm ver} &\ = \frac{E_F}{C_{\rm hfs}}\, (-4\pi i)
   \int \frac{d\bfp_1}{(2\pi)^3}\, \frac{d\bfp_2}{(2\pi)^3}\,
\nonumber \\ & \times
     \overline{\psi}_a(\bfp_1)\,
          \frac{\left[ \bfq \times {\bm \Gamma}_R(p_1,p_2)\right]_z}{\bfq^2}\,
       \psi_a(\bfp_2)\,,
\end{align}
where
$\bfq = \bfp_1-\bfp_2$, $p_1$ and $p_2$ are 4-vectors with the fixed time
component $p_1 = (\vare_a,\bfp_1)$, $p_2 = (\vare_a,\bfp_2)$, and ${\bm
 \Gamma}_R$ is the renormalized one-loop vertex operator. 
For evaluating the integrals over the angular variables in Eq.~(\ref{e6}), it is
convenient to employ the following representation of the
vertex operator sandwiched between the Dirac wave functions
\begin{align} \label{e7}
\overline{\psi}_a(\bfp_1)&\!\ {\bm\Gamma}_R(p_1,p_2)\,
        \psi_b(\bfp_2)
         = \frac{\alpha}{4\pi} 
\nonumber \\ \times  &\
\left[
    {\cal R}_1 \chi^{\dag}_{\kappa_a \mu_a}(\hat{\bfp}_1)
        {\bsigma} \chi_{-\kappa_a\mu_a}(\hat{\bfp}_2)
        \right. \nonumber \\ &
   +{\cal R}_2 \chi^{\dag}_{-\kappa_a \mu_a}(\hat{\bfp}_1)
        {\bsigma} \chi_{\kappa_a\mu_a}(\hat{\bfp}_2)
         \nonumber \\
&+ ({\cal R}_3 \bfp_1+ {\cal R}_4 \bfp_2)
         \chi^{\dag}_{\kappa_a \mu_a}(\hat{\bfp}_1)
         \chi_{\kappa_a\mu_a}(\hat{\bfp}_2)  \nonumber \\
&  + \left. ({\cal R}_5 \bfp_1 +{\cal R}_6 \bfp_2)
         \chi^{\dag}_{-\kappa_a \mu_a}(\hat{\bfp}_1)
         \chi_{-\kappa_a\mu_a}(\hat{\bfp}_2) \right] \,,
\end{align}
where the scalar functions ${\cal R}_i \equiv {\cal R}_i(p_{1r},p_{2r},q_{r})$ are
given by Eqs.~(A7)---(A12) of Ref.~\cite{yerokhin:99:sescr}, and 
$\hp_i \equiv \bfp_i/|\bfp_i|$, 
$p_{ir} =|\bfp_i|$, and $q_{r} = |\bfq|$. The dependence of the
integrand of Eq.~(\ref{e6})  on the angular variables can now be 
parameterized in terms of the basic
angular integrals $ K_i$ introduced and evaluated in
Appendix~\ref{app:1}. 
The result is
\begin{align} \label{e8}
\Delta E^{(0)}_{\rm ver} &\ = \frac{E_F}{C_{\rm hfs}}\, 
   \frac{\alpha}{48\pi^5}\,
      \int_0^{\infty}dp_{1r}\,dp_{2r}\int_{|p_{1r}-p_{2r}|}^{p_{1r}+p_{2r}}dq_r\,
           \frac{p_{1r}p_{2r}}{q_r}\,
\nonumber \\ & \times
     \Bigl\{ [-p_{1r}K_1(\kappa_a)+p_{2r}K_1^{\prime}(\kappa_a)] {\cal R}_1 
\nonumber \\ & 
           + [-p_{1r}K_1(-\kappa_a)+p_{2r}K_1^{\prime}(-\kappa_a)] {\cal R}_2  
\nonumber \\ & 
        - p_{1r}p_{2r}K_2(\kappa_a)\,({\cal R}_3+{\cal R}_4)
\nonumber \\ & 
        - p_{1r}p_{2r}K_2(-\kappa_a)\,({\cal R}_5+{\cal R}_6)\Bigr\}\,.
\end{align}
The above equation was used for the numerical evaluation. It contains four
integrations (the fourth one, over the
Feynman parameter, is implicit in the definition of the functions $R_i$). All
the integration were performed using Gauss-Legendre quadratures,
after appropriate substitutions in the integration variables. We
note that the integration variables $p_{1r}$, $p_{2r}$, and
$q_r$ resemble the well-known perimetric coordinates
\cite{pekeris:58}, in the sence that they 
weaken the (integrable) Coulomb singularity of the
integrand at $q_r = 0$.

%
%
\subsection{One-potential vertex part}
\label{onepotver}

The one-potential hfs vertex part $\Delta E_{\rm ver}^{(1)}$ is given by
\begin{align} \label{f1}
\Delta E_{\rm ver}^{(1)} &\ =  \frac{E_F}{C_{\rm hfs}}\,8\pi i Z \alpha^2\,
   \int \frac{d\bfp\, d\bfp\pr\, d\bfp\ppr}{(2\pi)^9}\,
     \overline{\psi}_a(\bfp)\,
\nonumber \\ & \times
          \frac{\left[ \bfp\ppr \times {\bm
                \Lambda}(p,p\pr,p\ppr)\right]_z}
      {(\bfq-\bfp\ppr)^2\, {\bfp\ppr}^2}\,
       \psi_a(\bfp\pr)\,,
\end{align}
where $\bfq = \bfp-\bfp\pr$ is the total momentum transfer
(final minus initial) for the electron vertex function $\bm \Lambda$, 
and the time component of the 4-vectors is fixed by $p_0 = p\pr_0 =
\vare_a$ and $p\ppr_0 = 0$. The 4-point vertex function $\Lambda$ is given by
\begin{align} \label{f2}
 &\ \Lambda_j(p,p\pr,p\ppr) = \frac{16\pi^2}{i} \int \frac{d^4 k}{(2\pi)^4}\,
\frac{1}{k^2}\,
\\ & \times
  \frac{\gamma_{\sigma} (\cross{p}-\cross{k}+m)\gamma_0
    (\cross{p}-\cross{k}-\cross{p}\ppr+m)
           \gamma_j (\cross{p}\pr-\cross{k}+m) \gamma^{\sigma}}
  {[(p-k)^2-m^2] [(p-k-p\ppr)^2-m^2] [(p\pr-k)^2-m^2]}\,.
\nonumber 
\end{align}
The evaluation of $\Delta E_{\rm ver}^{(1)}$ is performed by using the standard
technique for the evaluation of Feynman diagrams (for a short summary of the
relevant formulas, see Appendix D of Ref.~\cite{yerokhin:03:epjd}). First, we
use three Feynman parameters in order to join the 4 factors in the
denominator of the integrand in Eq.~(\ref{f2}). Denoting the numerator as
$N_j(k)$, we obtain
\begin{align} \label{f3}
 \Lambda_j(p,p\pr,p\ppr) &\ = \int dx\,dy\,dz\, 6x^2y\,
\nonumber \\ & \times
\frac{16\pi^2}{i} \int \frac{d^4 k}{(2\pi)^4}\, 
\frac{N_j(k)}{[(k-xb)^2-x\Delta]^4}\,,
\end{align}
where $x$, $y$, and $z$ are the Feynman parameters (here and below it is
assumed that all integrals over the Feynman parameters extend from 0 to 1).
We denote $b = (1-y)p+yp\pr+yz p\ppr$, and $\Delta = x b^2 +
m^2-(1-y)p^2-y{p\pr}^2-yz({p\ppr}^2+2p\pr \cdot p\ppr)$. 

Next, we shift the
integration variable $k\to k+xb$ and perform the integration over $k$. The
result is
\begin{align} \label{f4}
 \Lambda_j(p,p\pr,p\ppr) = \int dx\,dy\,dz\, y\,
\left[
\frac{N_j(xb)}{\Delta^2}- \frac{x N_{2,j}^{\mu\nu} g_{\mu\nu}}{2 \Delta}
\right]\,,
\end{align}
where $N_{2,j}$ is defined so that $N_{2,j}^{\mu\nu}k_{\mu}k_{\nu}$ is the  quadratic
in $k$ part of $N_j(k)$. We note that, after shifting the integration variable, only
even powers of $k$ yield a non-zero contribution to the integral (i.e.,
the terms proportional $k_{\mu}$ and $k_{\mu}k_{\nu}k_{\rho}$ vanish).

Next, the integration over $p\ppr$ is carried out. We introduce the
function $\Xi_{ij}$ by
\begin{align} \label{f5}
 \Xi_{ij}(p,p\pr) &\ \equiv \int \frac{d\bfp\ppr}{(2\pi)^3}\, \frac{p\ppr_i\,
   \Lambda_j(p,p\pr,p\ppr)} {{\bfp\ppr}^2 (\bfq-\bfp\ppr)^2}
\nonumber \\ &
= \int dx\,dy\,dz\, y\,
  \int \frac{d\bfp\ppr}{(2\pi)^3}\,
    \frac1{{\bfp\ppr}^2 (\bfq-\bfp\ppr)^2}
\nonumber \\ & \times
\left[
\frac{p\ppr_i\,N_{0,j}(p\ppr)}{\Delta^2}- 2x\,
\frac{p\ppr_i\, N_{2,j}(p\ppr)}{\Delta} \right]\,,
\end{align}
where $N_{0,j}(p\ppr)\equiv N_j(xb)$ and 
$N_{2,j}(p\ppr) \equiv N_{2_j}^{\mu\nu} g_{\mu\nu}/4$.
The integral over $\bfp\ppr$ in Eq.~(\ref{f5}) can be expressed in terms of
the Lewis integral \cite{lewis:56}. However, we prefer to perform this integration 
straightforwardly by merging denominators using Feynman parametrization. In
this way, we end up with an additional integration to be performed numerically, but
the structure of the expressions involved becomes somewhat simpler. 

Let us illustrate the further evaluation by considering the contribution induced by the first term 
in the square brackets in Eq.~(\ref{f5}), which will be denoted by 
$\Xi_{0,{ij}}$. We merge the denominators by introducing two more Feynman
parameters,
\begin{align} \label{f6}
\frac1{{\bfp\ppr}^2 (\bfq-\bfp\ppr)^2\,\Delta^2} = 
   \int du\, dt\, \frac{6u^2t}{(wyz)^2}
        \frac{1}{[(\bfp\ppr-u \bfc)^2+ u\Omega]^4}\,,
\end{align}
where $w = 1-xyz$, 
\begin{equation}
\bfc = \frac{t}{w}[x(1-y)\bfp-(1-xy)\bfp\pr]+(1-t)\bfq\,,
\end{equation}
and
\begin{align}
\Omega &\ = -u\bfc^2 + (1-t)\bfq^2+ \frac{t}{wyz}
\nonumber \\ & \times
   \left\{ x[(1-y)p+y p\pr]^2+ m^2 -(1-y)p^2-y{p\pr}^2\right\}\,.
\end{align}
Substituting Eq.~(\ref{f6}) into Eq.~(\ref{f5}) and shifting the integration
variable, we get
\begin{align} \label{f7}
 \Xi_{0,{ij}}(p,p\pr) = \int d_F \frac{6u^2t}{yw^2z^2}\,
   \int \frac{d\bfp\ppr}{(2\pi)^3}
     \frac{M_{ij}}{({\bfp\ppr}^2+u\Omega)^4}\,,
\end{align}
where $d_F \equiv dx\,dy\,dz\,du\,dt$ and 
\begin{align}
M_{ij} = &\ (\bfp\ppr_i+u\bfc_i)\, N_{0,j}(\bfp\ppr+u\bfc)
 \nonumber \\ 
\equiv  &\ M_{0,{ij}}+ M_{1,{ij}}^k p\ppr_k
 + M_{2,{ij}}^{kl} p\ppr_k p\ppr_l
 \nonumber \\  &\
 + M_{3,{ij}}^{klm} p\ppr_k p\ppr_l p\ppr_m
 + M_{4,{ij}}^{klmn} p\ppr_k p\ppr_l p\ppr_m p\ppr_n\,.
\end{align}
The above equation defines the $M$ functions as the coefficients
from the expanded form of the expression 
$(\bfp\ppr_i+u\bfc_i)\, N_{0,j}(\bfp\ppr+u\bfc)$. 
Performing the integration over $\bfp\ppr$ in Eq.~\eqref{f7}, we obtain
\begin{align} \label{f8}
 \Xi_{0,{ij}}(p,p\pr) &\ = \frac1{32\pi}
\int d_F \frac{u^2t}{y \, w^2 \, z^2}\,
   \left\{ \frac{3M_{0,{ij}}}{(u\Omega)^{5/2}} 
       \right.
\nonumber \\ 
&   \left. 
  + \frac{M_{2,{ij}}^{kk}}{(u\Omega)^{3/2}}
  +
  \frac{M_{4,{ij}}^{kkll}+M_{4,{ij}}^{klkl}+M_{4,{ij}}^{kllk}}{(u\Omega)^{1/2}}
   \right\}\,.
\end{align}
Because $\Omega$ is linear in $u$, the integral over $u$ is elementary and can
be expressed in terms of logarithms. The four remaining integrations over the
Feynman parameters remain to be evaluated numerically. To complete the
evaluation of $\Xi_{0,{ij}}$, one needs to obtain explicit expressions for the
numerators $M_{l,{ij}}$ and to bring them to the standard form. Under
``the standard form'' we understand a linear combination of independent matrix
structures, see below. This is the most tedious part of the calculation 
since the expressions involved are very lenghty. Usage of 
symbolic computation packages is indispensable in this case.

Having obtained an expression for $\Xi_{ij}$, we write the correction to
the hfs as
\begin{align} \label{f9}
\Delta E_{\rm ver}^{(1)} &\ =  \frac{E_F}{C_{\rm hfs}}\,8\pi i Z \alpha^2\,
   \int \frac{d\bfp\, d\bfp\pr}{(2\pi)^6}\,
     \overline{\psi}_a(\bfp)\,
\nonumber \\ & \times
\Xi(p_r,p\pr_r,q_r;X_1,\ldots,X_{32})\,
       \psi_a(\bfp\pr)\,,
\end{align} 
where we used the notation $\Xi \equiv \epsilon_{0ij}\, \Xi_{ij}$ with
$\epsilon_{ijk}$ denoting the Levi-Civita symbol. In Eq.~(\ref{f9}),
we indicate
explicitly the dependence of $\Xi$ on 32 basic matrix structures $X_i$. The
main four of these are:  $[\bfp \times \bgamma]_z$, $[\bfp\pr \times
\bgamma]_z$, $[\bfp \times \bfp\pr]_z$, and $[\bgamma \times \bgamma]_z$. The
rest is obtained by multiplying each of them by $\cross{p}$, $\cross{p}\pr$,
$\cross{p}\, \cross{p}\pr$, $\gamma^0$, $\cross{p}\gamma^0$,
$\gamma^0\cross{p}\pr$, and $\cross{p} \gamma^0 \cross{p}\pr$. 

In order to perform the integration over all angular variables in
Eq.~(\ref{f9}) except for $\xi = \hp\cdot\hp\pr$, we define the angular
integrals $Y_i$ that correspond to the basic matrices $X_i$ by
\begin{align} \label{f10}
\int d\hp\,d\hp\pr\, &\
     \overline{\psi}_a(\bfp)\, X_i\, F(p_r,p\pr_r,q_r)\,
       \psi_a(\bfp\pr) 
\\ &
= \int_{-1}^1d\xi\, Y_i\, F(p_r,p\pr_r,q_r)\,,
\qquad i=1,\dots,32 \,.
\nonumber 
\end{align}
where $F$ is an arbitrary function. All $Y_i$ may be expressed in
terms of the elementary angular integrals listed in Appendix~\ref{app:1}.

Using the angular integrals $Y_i$, we can
write the final expression for the one-potential vertex term suitable for a
numerical evaluation,
\begin{align} \label{f11}
\Delta E_{\rm ver}^{(1)} &\ =  \frac{E_F}{C_{\rm hfs}}\,
  \frac{\alpha(\Za)}{6\pi^4}\, \int_{0}^{\infty}dp_r\, dp\pr_r\,
\nonumber \\ & \times
   \int_{|p_{r}-p_{r}\pr|}^{p_{r}+p_{r}\pr}dq_r\,
           p_r p\pr_r q_r\,
\Xi(p_r,p\pr_r,q_r;Y_1,\ldots,Y_{32})\,.
\end{align} 
Altogether, Eq.~(\ref{f11}) contains 7 integrations to be performed
numerically, 3 of them being written explicitly 
and 4 Feynman-parameter
integrations contained in the definition of the function $\Xi$. 
The numerical evaluation was performed using Gauss-Legendre
quadratures for all integrations. In order to prevent losses of accuracy due
to numerical cancellations, we used quadruple-precision arithmetic
(accurate to roughly 32 decimals) in a
small part of the code, which was identified to be numerically unstable. The
evaluation was rather time-consuming (about a month of processor time for each
value of $Z$ and each state) and was performed with the help of the parallel
computational environment at MPI Heidelberg.

The one-potential vertex part has been crucial to our calculation, and so 
it may be appropriate to summarize once more 
the basic steps in its evaluation:
First of all, let us recall that our ``one-potential vertex part'' 
actually involves two vertices inside the loop, one 
being a Coulomb vertex and the other being a magnetic vertex
(coupling to the external field). Therefore, there are three
fermion propagators inside the loop and one photon
propagator, necessitating the introduction of three
Feynman parameters to join denominators. The incoming Coulomb momentum and the 
exchanged momentum with the external field entail two
further Feynman parameters, one of which is integrated out 
analytically. In addition to the four remaining Feynman parameters,
we have two radial integrations over the absolute values of  the initial 
($\bfp\pr$) and final ($\bfp$) electron momenta, and an integration over the direction 
cosine $\xi$ (transformed by a
change of variable to an integration over $q_r=|\bfp-\bfp\pr|$).
The three additional integrations account for the 
resulting seven-dimensional integral. In the corresponding 
calculation in free QED, one could hope to carry out the radial integrations
analytically, because the incoming and outgoing fermions 
are on the mass shell and described by plane waves. 
Here, however, the bound states are being off the mass shell and
have a much more complicated structure, so that the radial integrations have
to be evaluated numerically. The separate calculation
of the full one-potential vertex part as described in the 
current section leads to a numerically
favourable scheme, because this part 
can be then subtracted from the integrand
of the remaining nonperturbative vertex contribution, thereby leading to 
a drastic improvement in the convergence of the resulting partial-wave
expansion (see Table~\ref{tab:ver2} below). 

%
%
\subsection{Many-potential vertex part}
\label{manypotver}

The general expression for the many-potential vertex part 
$\Delta E_{\rm ver}^{(2+)}$ is obtained from Eq.~(\ref{eq2}) by applying the
appropriate set of subtractions in the electron propagators. The required
subtractions are given by 
\begin{align} \label{g1}
G\, \delta V\, G &\ \to  G\, \delta V\, G -
 G^{(a)}\, \delta V\, G^{(a)} \\ &
-  G^{(0)}\, \delta V\, G^{(0)}
-  G^{(0)}\, \delta V\, G^{(1)} -  G^{(1)}\, \delta V\, G^{(0)}\,,
\nonumber 
\end{align}
where $G$ denotes the bound-electron propagator, $G^{(0)}$ is the
free-electron proparator, $G^{(1)}$ is the electron propagator with one
interaction with the binding Coulomb field, and $G^{(a)}$ is the
reference-state part of the bound-electron propagator. 
This subtraction takes into account all terms which have been
calculated separately using different approaches, as described above.

In order to perform a numerical evaluation of $\Delta E_{\rm  ver}^{(2+)}$, it
is convenient to rotate the integration contour of the photon energy $\omega$ from
$(-\infty,\infty)$ to be parallel to the imaginary axis of the $\omega$ complex
plane. In this work, we define a deformed 
$\omega$ integration countour $C_{LH}$ 
consisting of two parts, a low-energy part $C_L$ and a high-energy part $C_H$.
The low-energy part contains the interval $\omega \in (\Delta-i0,-i0)$ on the lower bank
of the cut of the photon propagator and the interval $(i0,\Delta+i0)$ on the
upper bank of the cut, with $\Delta = \Za\, \vare_a$. The high-energy part
consists of two intervals, $(\Delta+i0,\Delta+i\infty)$ and
$(\Delta-i0,\Delta-i\infty)$. The contour $C_{LH}$ defined in this way differs
from the one used by P.~J.~Mohr \cite{mohr:74:a} only by the choice of the
separation point $\Delta$ (the value $\Delta = \vare_a$ instead of $\Delta =
\Za\, \vare_a$ was employed in Ref.~\cite{mohr:74:a}). 

The high-energy part of $\Delta E_{\rm  ver}^{(2+)}$ is given by
\begin{align} \label{g2}
\Delta E_{{\rm ver},H}^{(2+)} &\ = -\frac{1}{\pi}\,{\rm Re}\int_{0}^{\infty} d\omega\,
  \nonumber \\
& \times   \sum_{n_1n_2} \Biggl[
    \frac{\lbr n_1|\delta V|n_2\rbr\,
            \lbr an_2| I(\Delta+i\omega)|n_1a\rbr}
                   {(\vare_a-\Delta-i\omega-\vare_{n_1})
                   (\vare_a-\Delta-i\omega-\vare_{n_2})}
\nonumber \\ &
-\ \mbox{\rm subtractions} \Biggr]
\,,
\end{align}
where the subtractions are given by Eq.~(\ref{g1}).
The low-energy part of $\Delta E_{\rm  ver}^{(2+)}$ needs a careful treatment
because of single and double poles situated near
the contour $C_L$, which are due to virtual bound states of lower
energy than the reference state. The single poles can be integrated via 
a Cauchy principal value prescription, and the double poles can be converted
to single poles via an integration by parts. We thus write the
low-energy part of $\Delta E_{\rm  ver}^{(2+)}$ as
\begin{align} \label{g3}
\Delta E_{{\rm ver},L}^{(2+)} &\ = -\frac{1}{\pi}\,P\int_{0}^{\Delta} d\omega\,
  \nonumber \\
& \times   \Biggl[
\sum_{\scriptsize \begin{array}{c} n_1 \, n_2\\
{\rm not}\, 0 <\vare_{n_1}=\vare_{n_2}<\vare_a \\
\end{array}}
    \frac{F_{n_1n_2}(\omega)}
                   {(\vare_a-\omega-\vare_{n_1})
                   (\vare_a-\omega-\vare_{n_2})}
\nonumber \\ 
& - \sum_{0 <\vare_n<\vare_a} 
    \frac{F\pr_{nn}(\omega)}
                   {\vare_a-\omega-\vare_{n}}
-\ \mbox{\rm subtractions} \Biggr]
\nonumber \\ &
-\frac{1}{\pi}\,
 \sum_{0 <\vare_n<\vare_a} 
    \frac{F_{nn}(\Delta)}
                   {\vare_a-\Delta-\vare_{n}}
\,,
\end{align}
where 
\begin{align} \label{g4}
F_{n_1n_2}(\omega) =   \lbr n_1|\delta V|n_2\rbr\,
\lbr an_2| {\rm Im}\left[I(\omega)\right]|n_1a\rbr\,,
\end{align} 
the prime denotes the derivative over $\omega$, and $P$ denotes the Cauchy
principal value of the integral. In Eq.~(\ref{g3}), all 
terms that induce double poles on the interval $\omega \in (0,\Delta)$
(i.e., intermediate states with $0<\vare_{n_1} = \vare_{n_2} < \vare_a$) have
been integrated by parts. We recall that the term with $\vare_{n_1} = \vare_{n_2} = \vare_a$
is removed by the $G^{(a)}$ part of the subtraction (\ref{g1}). 

The need to evaluate the principal 
value of the integral over $\omega$ complicates the numerical
calculation of the low-energy part. 
In the case when there is a single pole only 
(which takes place for the
$2s$ and $2p_{1/2}$ reference state), the problem is most
easily solved by employing a numerical quadrature symmeric around the position of the
pole. In the general case with more than one singularity 
to be treated, this approach is not effective. A better
way is to introduce subtractions in the integrand that remove the
singularities at the poles and to evaluate the principal value of the integral
of the subtracted terms analytically. We introduce the subtractions by
observing that the following difference does not have any singularities on the
interval $\omega \in (0,\Delta)$,
\begin{align} \label{g5}
& \sum_{\scriptsize \begin{array}{c} n_1 \, n_2\\
{\rm not}\, 0<\vare_{n_1}=\vare_{n_2}\leq\vare_a \\
\end{array}}
 \frac{ \lbr n_1|\delta V|n_2\rbr \lbr an_2| {\rm Im}\left[I(\omega)\right]|n_1a\rbr}
                   {(\vare_a-\omega-\vare_{n_1})
                   (\vare_a-\omega-\vare_{n_2})}
\nonumber \\
& \qquad - \sum_{0<\vare_{n_1}<\vare_a} \frac{\lbr a \delta n_1| 
 {\rm Im}\left[I(\vare_a-\vare_{n_1})\right]|n_1 a\rbr}
        {\vare_a-\omega-\vare_{n_1}}
\nonumber \\
& \qquad - \sum_{0<\vare_{n_2}<\vare_a} \frac{\lbr an_2| {\rm
     Im}\left[I(\vare_a-\vare_{n_2})\right]|\delta n_2a\rbr}
{\vare_a-\omega-\vare_{n_2}}\,,
\end{align} 
where 
\begin{align} \label{g6}
| \delta n_1\rbr = \sum_{n_2\ne n_1} 
 \frac{|n_2\rbr\lbr n_2|\delta V|n_1\rbr}{\vare_{n_1}-\vare_{n_2}}\,.
\end{align} 
We note that the terms with
$(\vare_{n_1}=\vare_a,\vare_{n_2}\neq\vare_a)$ and
$(\vare_{n_1}\neq\vare_a,\vare_{n_2}=\vare_a)$ present in Eq.~(\ref{g5}) do not induce any
singularitites because ${\rm Im}[I(0)]=0$.
The perturbed wave function $|\delta n_1\rbr$ is known analytically (for the
hfs perturbing potential, both the
diagonal and the non-diagonal in $\kappa$ parts; for the $g$-factor perturbing
potential, only the diagonal in $\kappa$ part) from 
the generalized virial relations for the Dirac equation 
\cite{shabaev:91:jpb,shabaev:03:PSAS}. 

In order to complete our discussion of the evaluation of the many-potential
vertex part, we present the explicit expression for it after the integration
over the angular variables. This expression reads
\begin{align}
\Delta E_{\rm ver}^{(2+)} &\ = \frac{E_F}{C_{\rm hfs}}\, \frac{i\alpha}{2\pi}
 \int_{C_{LH}} d\omega\,
  \sum_{n_1n_2L} \Biggl[ \SixJ{j_1}{j_2}{1}{j_a}{j_a}{L}\,
  \nonumber \\ & \times   
    \frac{
    P(n_1,n_2)\, R_L(\omega,an_2n_1a)}
          {(\vare_a-\omega-\vare_{n_1})
                   (\vare_a-\omega-\vare_{n_2})}
  \nonumber \\
& - \mbox{\rm subtractions} \Biggr]
\,,
\end{align}
where $R_L$ is a relativistic generalization of the 
Slater radial integral, whose explicit expression is given in
Ref.~\cite{yerokhin:99:pra}. $P(n_1,n_2)$ is given by
\begin{align}
P(n_1,n_2) &\ = (-1)^{j_a+1/2}\, C^{10}_{j_a -1/2, j_a 1/2}\, 
  \nonumber \\ & \times
       \frac{\kappa_1+\kappa_2}{\sqrt{3}}\, 
\lbr -\kappa_2||C^{(1)}||\kappa_1\rbr\,
          R_{-2}(n_1,n_2)\,,
\end{align}
where $C_{j_1m_1,j_2m_2}^{jm}$ is the Clebsch-Gordan coefficient, 
$C^{(l)}_m = \sqrt{4\pi/(2l+1)}\,Y_{lm}$ is a normalized spherical
harmonic, and
\begin{equation}
R_{-2}(n_1,n_2) = \int_0^{\infty}dr\,
  \bigl[g_{n_1}(r)f_{n_2}(r)+f_{n_1}(r)g_{n_2}(r)\bigr]\,.
\end{equation}

The numerical evaluation of the many-potential vertex contribution is the most
difficult part of the calculation. The key feature that limits the
accuracy achievable in a numerical calculation is the convergence of the
partial-wave expansion. We recall that 
the many-potential vertex contribution $\Delta E_{\rm ver}^{(2+)}$ 
contains two and more Coulomb interactions (and a magnetic interaction) inside the self-energy loop. 
The convergence of its partial-wave expansion is much better than that for 
the vertex contribution with just one Coulomb interaction. In order to
illustrate this point, Table~\ref{tab:ver2} presents a comparison of the
partial-wave expansion of the vertex contribution with two and more Coulomb
interactions, $\Delta E_{\rm ver}^{(2+)}$, and of that with one and more Coulomb
interactions, $\Delta E_{\rm ver}^{(1+)}$. It can be seen that the subtraction
of the one-potential vertex contribution improves the numerical accuracy by
about 5 orders of magnitude. 

%
%
\begin{table*}[thb]
\caption{Comparison of the convergence of the partial-wave expansion for the 
corrections $\Delta E_{\rm ver}^{(1+)}$ and $\Delta E_{\rm ver}^{(2+)}$ 
for the hfs of the $1S$ state of atomic hydrogen
($Z=1$), in units $\Delta E/[\alpha/\pi\, E_F]$. 
$S(\kappa_{\rm max})$ is the sum of all partial contributions with
$|\kappa| \leq \kappa_{\rm max}$, and the convergence is measured
as $\kappa_{\rm max}$ is increased.  $\delta S$ is the increment. 
For $\Delta E_{\rm ver}^{(2+)}$, at the same value of $\kappa_{\rm max}$,
the apparent convergence gives us
roughly five more decimals as compared to $\Delta E_{\rm ver}^{(1+)}$.
 \label{tab:ver2} }
\begin{ruledtabular}
\begin{tabular}{r....}
& \multicolumn{2}{c}{$\Delta E_{\rm ver}^{(1+)}$}  &
\multicolumn{2}{c}{$\Delta E_{\rm ver}^{(2+)}$} \\[2ex]
$\kappa_{\rm max}$  
&\multicolumn{1}{c}{$\delta S$} & \multicolumn{1}{c}{$S(\kappa_{\rm max})$}  
&\multicolumn{1}{c}{$\delta S$} & \multicolumn{1}{c}{$S(\kappa_{\rm max})$}  \\
\hline\\[-9pt]
\rule[-2mm]{0mm}{6mm}
   3    &   1.58x0875 &   1.58x0875    &  3.36124x192571    &      3.36124x192571 \\
\rule[-2mm]{0mm}{6mm}
   7    &  -0.00x2317 &   1.57x8558    & -0.00000x693015    &      3.36123x499556 \\
\rule[-2mm]{0mm}{6mm}
  15    &  -0.00x0660 &   1.57x7898    & -0.00000x079070    &      3.36123x420487 \\
\rule[-2mm]{0mm}{6mm}
  30    &  -0.00x0183 &   1.57x7715    & -0.00000x009848    &      3.36123x410639 \\
\rule[-2mm]{0mm}{6mm}
  60    &  -0.00x0049 &   1.57x7666    & -0.00000x001246    &      3.36123x409393 \\
\rule[-2mm]{0mm}{6mm}
 120    &  -0.00x0010 &   1.57x7656    & -0.00000x000122    &      3.36123x409270 \\
\rule[-2mm]{0mm}{6mm}
 extrap.& -0.00x0015(15)&1.57x7641(15) & -0.00000x000041(44)&     3.36123x409229(44) \\
%
%
\end{tabular}
\end{ruledtabular}
\end{table*}

The partial-wave expansion was cut off at the maximum value of 
$|\kappa_{\rm max}| = 120$. Quadruple-precision 
arithmetics was required for the
Dirac Green function in order to control the numerical 
accuracy at the required level. This computation was performed with a
quadruple-precision generalization of the code for the Dirac Green function
developed in Refs.~\cite{yerokhin:99:pra,yerokhin:01:hfs}. It should be
mentioned that the evaluation of the high-energy part of 
$\Delta E_{\rm ver}^{(2+)}$ 
with the integration contour $C_{LH}$
requires the Dirac Green function with general complex values of the
energy argument. (This is in contrast to the approach used in 
Refs.~\cite{mohr:74:a,jentschura:99:prl,jentschura:01:pra}, where the
integration contour is chosen in such a way that only the real and purely
imaginary values of the energy argument are required.) The computation of the
Dirac Green function for general complex energies $\omega$ becomes numerically
unstable when $\kappa$ is large and ${\rm arg}(\omega)$ is close to
$\pi/4$. Because of this, we were not able to extend the partial-wave
summation beyond $|\kappa_{\rm  max}| = 120$. 

The general scheme of our evaluation is as
follows. We perform the summation over $\kappa$ directly in the integrand,
before any integrations. The summation is terminated when a suitable convergence
criterion is fulfilled or when the cutoff value $|\kappa_{\rm  max}|$ is
reached. In order to estimate the dependence of the final result on the cutoff 
parameter, results for several intermediate cutoffs are stored, each
consequent one being twice larger than the previous (see Table~\ref{tab:ver2}
for an illustration). The omitted tail of the expansion was estimated by using
the $\epsilon$ algorithm for Pad\'e approximation, and the uncertainty of the
extrapolation was taken about 50\%-200\% of the estimated tail. 

%
%
\subsection{Many-potential reducible part}
\label{manypotred}

According to Eq.~(\ref{eq8}), the reducible part of the 
SE correction involves the derivative of the SE 
operator, ``sandwiched'' in the reference state.
The zero-potential part of the reducible contribution
has already been discussed in Sec.~\ref{zeropotredver};
it involves the derivative of the free electron propagator 
with respect to the reference-state energy.
The reference-state contribution to the reducible 
part has been treated in Sec.~\ref{refstatecontrib},
together with the reference-state contribution
to the vertex term, thereby mutually cancelling the IR divergence
inherent to both reference-state contributions. 
The total reference-state contribution is summarized 
in Eq.~(\ref{d6}). Left is the many-potential reducible part,
\begin{align}
\label{ered1+}
& \Delta E^{(1+)}_{\rm red} = \lbr a|\delta V|a\rbr 
\nonumber\\
& \quad \times \lbr a| \gamma^0 \left.
    \frac{\partial}{\partial \vare} 
( \Sigma(\vare) - \Sigma^{(0)}(\vare) - \Sigma^{(a)}(\vare) )
\right|_{\vare = \vare_a} | a\rbr\,.
\end{align}
Here, $\Sigma^{(0)}(\vare)$ and $\Sigma^{(a)}(\vare)$
are obtained from Eq.~(\ref{eq6}) by a replacement of the
full Dirac--Coulomb Green function $G$ by the free Green 
function $G^{(0)}$ and by the reference-state part of the propagator $G^{(a)}$.
For the term with $G^{(a)}$, we have
\begin{eqnarray} 
 \Sigma^{(a)}(\vare,\bfx_1,\bfx_2) &=&
2\,i\alpha\,\gamma^0 \int_{-\infty}^{\infty} d\omega\,
      \alpha_{\mu}\,
\\ && \times
         G^{(a)}(\vare-\omega,\bfx_1,\bfx_2)\, \alpha_{\nu}\,
    D^{\mu\nu}(\omega,\bfx_{12})\,
         \,.
\nonumber
\end{eqnarray}
In coordinate space, a representation of $G^{(a)}$ reads
\begin{equation}
G^{(a)}(\vare-\omega,\bfx_1,\bfx_2) =
\sum_{\scriptsize \begin{array}{cc} n \\[-0.5ex]
{\vare_n = \vare_a} \end{array}}
\frac{\psi_n(\bfx_1) \, \psi_n^+(\bfx_2)}{\vare-\omega-\vare_a + i \, 0} \,,
\end{equation}
where we take into account all states with the same energy 
as the reference state, i.e., also the state with 
opposite parity but the same total angular momentum
as compared to the reference state (pairs of states with
the same $|\kappa|$ are energetically degenerate 
according to Dirac theory).

The evaluation of Eq.~(\ref{ered1+}) proceeds along the 
integration contour $C_M$ of P. J. Mohr~\cite{mohr:74:a}
for the (complex rather than real) photon energy. It is divided into a 
low-energy and a high-energy part.
The low-energy contour $C'_L$ comprises the interval $\omega \in (\vare_a-i0,-i0)$ below
the cut of the photon propagator and the interval $(i0,\vare_a+i0)$ on the
upper bank of the cut, with $\vare_a$ being the reference-state energy. 
The high-energy contour $C'_H$ again consists of two intervals, $(\vare_a+i0,\vare_a+i\infty)$ and
$(\vare_a-i0,\vare_a-i\infty)$. Because the low-energy part extends 
to comparatively high values of $|\omega|$, the 
radial integrand for each single value of $\kappa$ become
highly oscillatory. The behaviour of the integrand can only 
be improved if the full sum over intermediate angular
momenta is carried out before the radial integrations.
This is already evident from the model example
given in Eq.~(7.3) of Ref.~\cite{jentschura:99:cpc},
\begin{equation}
\label{ht1}
\frac{\exp\bigl( -r[1-\rho] \bigr)}{r[1-\rho]} \; = \; - \,
\sum_{|\kappa|=0}^{\infty} (2|\kappa|+1) \, j_{|\kappa|} (i\rho \, r) \, 
h^{(1)}_{|\kappa|} (i r) \,,
\end{equation}
where $j$ is a Bessel function and $h^{(1)}$ is a Hankel function of the 
first kind ($0 < \rho < 1$). The right-hand side of
Eq.~(\ref{ht1}) involves functions that are highly oscillatory
as a function of the radial variable $r$, but the left-hand side is a simple exponential.
This ``smoothing'' phenomenon after the summation over the 
intermediate angular momenta is crucial for the 
evaluation as it enhances the rate of convergence of the 
multi-dimensional SE integrals dramatically. The convergence of the 
sum over $|\kappa|$ can be further accelerated 
by the so-called CNC transformation~\cite{jentschura:99:cpc}.
With maximum values of $\kappa$ in excess of $10^6$ being handled at ease
using the CNC transformation, we are able to 
control the accuracy of the final evaluations.
The derivative of the Green function is calculated directly
using fourth-point and (alternatively, for verification) six-point
difference schemes. We choose suitable values of the parameters so that 
the Green function derivative is calculated to a relative accuracy of
$10^{-24}$. Additional modifications are necessary in the extreme
infrared region of photon energies; here the difference scheme is 
adjusted so that the boundaries of the integration region are not 
crossed and sufficient accuracy is retained.
A numerical subtraction of all singular terms due to 
lower-lying atomic states (e.g., the ground state) before 
doing any integrations over the photon energies and before 
evaluating the derivative of the Dirac propagator eliminates
a potential further source of numerical loss of significance
for the many-potential reducible part.

%
%
\subsection{Irreducible part}
\label{irreducible}

With reference to Eq.~(\ref{eq5}),
we recall that the irreducible part is given as
\begin{equation}
\Delta E_{\rm ir} = \lbr \delta a|\gamma^0
    \widetilde{\Sigma}(\vare_a) |a\rbr
              + \lbr a|\gamma^0 \widetilde{\Sigma}(\vare_a) |\delta a\rbr\,,
\end{equation}
with the renormalized SE operator $\widetilde{\Sigma}$ and
the perturbed wave function [see Eq.~(\ref{eq4})]

\begin{equation}
|\delta a\rbr = 
\sum_{\scriptsize \begin{array}{cc} n \\[-0.5ex]
{\vare_n \ne \vare_a} \end{array}}
\frac{|n\rbr \lbr n|\delta V|a\rbr}{\vare_a-\vare_n} \,.
\end{equation}
We only need the diagonal-in-$\kappa$ component of 
the perturbed wave function, because the 
SE operator is also diagonal in the total
angular momentum.

The evaluation of the irreducible part is carried out 
along the same contour $C_M$ that is used for the many-potential
reducible part. Within the high-energy part,
the Green function is divided 
into two parts. The first is 
a subtraction term which involves a free propagator and an approximate 
one-potential term~\cite{mohr:74:a}, which is obtained
from the full one-potential term by commuting the Coulomb potential
to the left of the electron propagators.
The second is the remainder term
which is the difference of the full and the approximate propagator.
The subtraction term contains all UV divergences 
of the irreducible part; these are cancelled against the mass counter term
$\delta m$. The subtraction term is evaluated in momentum 
space, in a noncovariant integration scheme adjusted for bound-state
calculations, where the spatial components of the photon
momentum are integrated out before the photon energy 
integration.

%
%
\begin{table}[thb]
\caption{Individual contributions to the SE correction to the hfs of the
$1S$ state of hydrogen, in units $\Delta E/[\alpha/\pi\, E_F]$. 
The specific contributions are discussed in 
Sec.~\ref{refstatecontrib} ($\Delta E^{(a)}$), 
Sec.~\ref{zeropotredver} ($\Delta E_{\rm red}^{(0)}$ and $\Delta E_{\rm ver}^{(0)}$),
Sec.~\ref{onepotver} ($\Delta E_{\rm ver}^{(1)}$),
Sec.~\ref{manypotver} ($\Delta E_{\rm ver}^{(2+)}$),
Sec.~\ref{manypotred} ($\Delta E_{\rm red}^{(1+)}$), and
Sec.~\ref{irreducible} ($\Delta E_{\rm ir}$).
\label{tab:breakdown} }
\begin{tabular}{l@{\hspace*{0.5cm}}.}
\hline
\hline
\rule[-2mm]{0mm}{6mm}
$\Delta E_{\rm ir}$         &  -0.0109x6549784\,(5) \\
\rule[-2mm]{0mm}{6mm}
$\Delta E_{\rm red}^{(0)}$  &   8.2895x6864683      \\
\rule[-2mm]{0mm}{6mm}
$\Delta E_{\rm red}^{(1+)}$ &  -3.8385x4412893\,(5) \\
\rule[-2mm]{0mm}{6mm}
$\Delta E_{\rm ver}^{(0)}$  &  -5.5795x8625925      \\
\rule[-2mm]{0mm}{6mm}
$\Delta E_{\rm ver}^{(1)}$  &  -1.7835x813412\,(16) \\
\rule[-2mm]{0mm}{6mm}
$\Delta E_{\rm ver}^{(2+)}$ &   3.3612x340923\,(4)  \\
\rule[-2mm]{0mm}{6mm}
$\Delta E^{(a)}$            &  -0.0000x2366906      \\
\rule[-2mm]{0mm}{6mm}
Total                       &   0.4381x018429\,(16) \\
\hline
\hline
\end{tabular}
\end{table}

%
%
\begin{table}[thb]
\caption{SE correction to the hfs of $nS$ states of hydrogen-like ions. $\delta E_{nS}
= \Delta E_{nS}/[\alpha/\pi\, E_F(nS)]$ and $F_{nS}$ is the higher-order remainder
defined by Eq.~(\ref{ee2}). 
 \label{tab:hfs:s} }
\begin{ruledtabular}
\begin{tabular}{cr..}
$1S$ & $Z$ 
&  \multicolumn{1}{c}{$\delta E_{1S}$} 
& \multicolumn{1}{c}{$F_{1S}(\Za)$}  \\
\hline\\[-9pt]
 & 1 &   0.43810x18429\,(16) &  -13.83x08\,(43) \\
 & 2 &   0.37346x7600\,(3) &  -14.11x70\,(9) \\
 & 3 &   0.30758x3838\,(4) &  -14.41x21\,(3) \\
 & 4 &   0.24100x5731\,(5) &  -14.69x62\,(2) \\
 & 5 &   0.17402x6212\,(7) &  -14.96x73\,(2) \\
 & 6 &   0.10681x5805\,(11)&  -15.22x64\,(1) \\
 & 7 &   0.03947x649\,(2)  &  -15.47x52\,(1) \\
 & 8 &  -0.02793x233\,(2)  &  -15.71x56\,(1) \\
 & 9 &  -0.09537x946\,(3)  &  -15.94x87\,(1) \\
 &10 &  -0.16285x352\,(4)  &  -16.17x62\,(1) \\
 &11 &  -0.23035x739\,(4)  &  -16.39x90\,(1) \\
 &12 &  -0.29790x470\,(4)  &  -16.61x81\,(1) \\
\hline
\hline\\[-9pt]
$2S$ & $Z$ 
& \multicolumn{1}{c}{$\delta E_{2S}$} 
& \multicolumn{1}{c}{$F_{2S}(\Za)$}  \\
\hline\\[-9pt]
 & 1 &   0.43869x2275\,(3) &   -6.12x05\,(85) \\
 & 2 &   0.37535x2042\,(3) &   -6.91x26\,(11) \\
 & 3 &   0.31120x3194\,(5) &   -7.58x33\,(5) \\
 & 4 &   0.24666x5425\,(7) &   -8.16x97\,(3) \\
 & 5 &   0.18193x8687\,(9) &   -8.70x69\,(2) \\
 & 6 &   0.11712x3392\,(13)&   -9.19x73\,(2) \\
 & 7 &   0.05226x463\,(2)  &   -9.65x46\,(1) \\
 & 8 &  -0.01262x594\,(2)  &  -10.08x59\,(1) \\
 & 9 &  -0.07755x851\,(3)  &  -10.49x64\,(1) \\
 &10 &  -0.14255x836\,(3)  &  -10.89x01\,(1) \\
 &11 &  -0.20766x159\,(3)  &  -11.27x01\,(1) \\
 &12 &  -0.27291x238\,(4)  &  -11.63x90\,(1) \\
\hline  
\hline\\[-9pt]
$3S$ & $Z$ 
& \multicolumn{1}{c}{$\delta E_{3S}$} 
& \multicolumn{1}{c}{$F_{3S}(\Za)$}  \\
\hline\\[-9pt] 
 & 1 &   0.43893x143\,(3) &   -1.72x7\,(69) \\
 & 2 &   0.37613x687\,(3) &   -2.61x38\,(90) \\
 & 3 &   0.31274x865\,(3) &   -3.35x30\,(27) \\
 & 4 &   0.24914x165\,(3) &   -3.99x94\,(12) \\
 & 5 &   0.18548x734\,(3) &   -4.58x06\,(7) \\
 & 6 &   0.12186x545\,(3) &   -5.11x33\,(4) \\
 & 7 &   0.05830x586\,(3) &   -5.60x92\,(2) \\
 & 8 &  -0.00519x144\,(3) &   -6.07x57\,(1) \\
 & 9 &  -0.06864x589\,(3) &   -6.51x89\,(1) \\
 &10 &  -0.13209x008\,(3) &   -6.94x30\,(1) \\
 &11 &  -0.19556x594\,(4) &   -7.35x15\,(1) \\
 &12 &  -0.25912x227\,(5) &   -7.74x73\,(1) \\
\end{tabular}
\end{ruledtabular}
\end{table}

%
%
\begin{table}[thb]
\caption{SE correction to the hfs of $2P_J$ states of hydrogen-like ions. $\delta E_{nP_J}
= \Delta E_{nP_J}/[\alpha/\pi\, E_F(nP_J)]$ and $G_{nP_J}$ is the higher-order remainder
defined by Eq.~(\ref{ee3}).
 \label{tab:hfs:p} }
\begin{ruledtabular}
\begin{tabular}{cr..}
$2P_{1/2}$ & $Z$ 
& \multicolumn{1}{c}{$\delta E_{2P_{1/2}}$} 
& \multicolumn{1}{c}{$G_{2P_{1/2}}(\Za)$}  \\
\hline\\[-9pt]
 & 1 &   0.24939x7018\,(5) &   -11.3233x21\,(86) \\
 &   &   0.2487\,(x5)^a    &        \\
 & 2 &   0.24801x6543\,(5) &    -9.3117x68\,(25) \\
 & 3 &   0.24608x7170\,(7) &    -8.1642x78\,(14) \\
 & 4 &   0.24371x9931\,(7) &    -7.3707x86\,(9) \\
 & 5 &   0.24098x5405\,(8) &    -6.7713x55\,(6) \\
 &   &   0.2397^ax         &\\
 & 6 &   0.23793x2761\,(8) &    -6.2946x96\,(4) \\
 & 7 &   0.23459x7810\,(8) &    -5.9027x68\,(3) \\
 & 8 &   0.23100x7222\,(8) &    -5.5728x57\,(2) \\
 & 9 &   0.22718x0996\,(8) &    -5.2903x09\,(2) \\
 &10 &   0.22313x4035\,(8) &    -5.0451x23\,(2) \\
 &   &   0.2202^ax         &\\
 &11 &   0.21887x7214\,(8) &    -4.8301x70\,(1) \\
 &12 &   0.21441x8110\,(9) &    -4.6401x91\,(1) \\
\hline
\hline\\[-9pt]
$2P_{3/2}$ & $Z$ 
& \multicolumn{1}{c}{$\delta E_{2P_{3/2}}$} 
& \multicolumn{1}{c}{$G_{2P_{3/2}}(\Za)$}  \\
\hline\\[-9pt]
 & 1 &  -0.12499x329\,(1) &     0.1260x9\,(18) \\
 &   &  -0.1254^bx&\\
 & 2 &  -0.12498x309\,(1) &     0.0794x05\,(55) \\
 & 3 &  -0.12498x458\,(2) &     0.0321x76\,(39) \\
 & 4 &  -0.12501x321\,(2) &    -0.0154x99\,(29) \\
 & 5 &  -0.12508x457\,(3) &    -0.0635x28\,(21) \\
 &   &  -0.1255^bx&\\
 & 6 &  -0.12521x458\,(3) &    -0.1119x33\,(16) \\
 & 7 &  -0.12541x922\,(3) &    -0.1606x63\,(12) \\
 & 8 &  -0.12571x467\,(3) &    -0.2096x98\,(9) \\
 & 9 &  -0.12611x728\,(3) &    -0.2590x28\,(7) \\
 &10 &  -0.12664x357\,(3) &    -0.3086x43\,(5) \\
 &   &  -0.1271^bx&\\
 &11 &  -0.12731x021\,(3) &    -0.3585x38\,(5) \\
 &12 &  -0.12813x408\,(3) &    -0.4087x11\,(4) \\
\end{tabular}
\end{ruledtabular}
$^a$ Ref.~\cite{sapirstein:06:hfs}\\
$^b$ Ref.~\cite{sapirstein:08:hfs}\\
\end{table}

%
%
\begin{figure}[thb]
\includegraphics[width=0.9\linewidth]{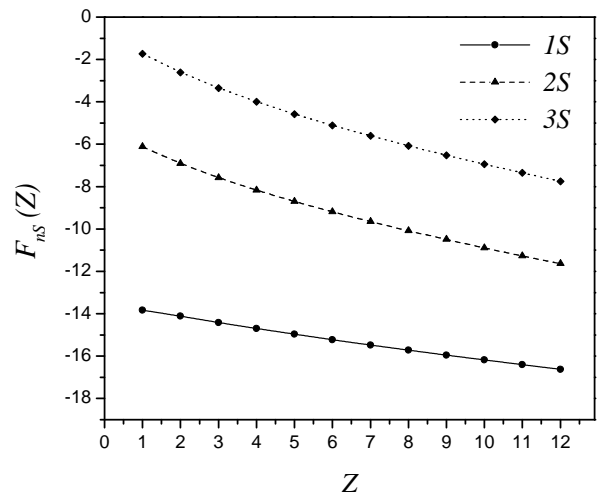}
\caption{The higher-order remainder $F_{nS}(Z\alpha)$ 
for the SE correction to the hfs of 
the $1S$, $2S$, and $3S$  states. \label{fig:hos}}
\end{figure}

%
%
\begin{figure}[thb]
\includegraphics[width=0.9\linewidth]{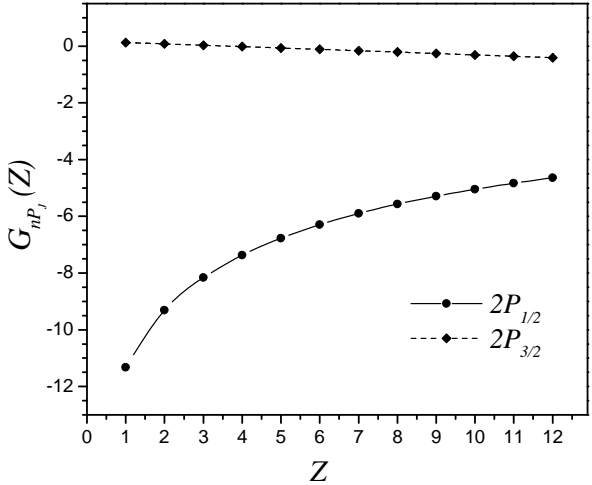}
\caption{The higher-order remainder $G_{nP_J}(Z\alpha)$ 
for the SE correction to the hfs of the
$2P_{1/2}$ and  $2P_{3/2}$ states. 
 \label{fig:hop}}
\end{figure}

%
%
\begin{table}
\caption{SE correction to the $g$-factor of $nS$ states of hydrogen-like ions,
 in ppm.  $H_{nS}$ is the higher-order remainder.
 \label{tab:gfact:s} }
\begin{ruledtabular}
\begin{tabular}{cr..}
$1S$ & $Z$ 
& \multicolumn{1}{c}{$\Delta g_{1S}$} 
& \multicolumn{1}{c}{\hskip0.5cm $H_{1S}(\Za)$}  \\
\hline\\[-9pt]
 & 1 &     2322.84x0230\,(2)   &     19x.\,(3.) \\
 & 2 &     2322.90x4037\,(4)   &     23x.3\,(2.6) \\
 & 3 &     2323.01x4295\,(8)   &     22x.88\,(70) \\
 & 4 &     2323.17x5525\,(14)  &     22x.58\,(29) \\
 & 5 &     2323.39x298\,(2)    &     22x.36\,(15) \\
 & 6 &     2323.67x243\,(3)    &     22x.18\,(9) \\
 & 7 &     2324.02x001\,(5)    &     22x.02\,(6) \\
 & 8 &     2324.44x213\,(7)    &     21x.87\,(4) \\
 & 9 &     2324.94x538\,(8)    &     21x.71\,(3) \\
 &10 &     2325.53x651\,(10)   &     21x.57\,(2) \\
 &11 &     2326.22x235\,(13)   &     21x.42\,(2) \\
 &12 &     2327.00x983\,(12)   &     21x.28\,(1) \\
\hline
\hline\\[-9pt]
$2S$ & $Z$ 
& \multicolumn{1}{c}{$\Delta g_{2S}$} 
& \multicolumn{1}{c}{\hskip0.5cm $H_{2S}(\Za)$}  \\
\hline\\[-9pt]
 & 1 &     2322.82x4624\,(2)   &     \\
 & 2 &     2322.84x0323\,(8)   &     \\
 & 3 &     2322.86x696\,(2)    &     19x.\,(11.) \\
 & 4 &     2322.90x509\,(3)    &     22x.1\,(4.7) \\
 & 5 &     2322.95x533\,(5)    &     22x.5\,(2.4) \\
 & 6 &     2323.01x836\,(6)    &     22x.5\,(1.3) \\
 & 7 &     2323.09x490\,(8)    &     22x.43\,(74) \\
 & 8 &     2323.18x571\,(8)    &     22x.31\,(42) \\
 & 9 &     2323.29x154\,(9)    &     22x.18\,(24) \\
 &10 &     2323.41x319\,(9)    &     22x.05\,(15) \\
 &11 &     2323.55x142\,(11)   &     21x.92\,(11) \\
 &12 &     2323.70x704\,(12)   &     21x.79\,(8) \\
\hline
\hline\\[-9pt]
$3S$ & $Z$ 
& \multicolumn{1}{c}{$\Delta g_{3S}$} 
& \multicolumn{1}{c}{\hskip0.5cm $H_{3S}(\Za)$}  \\
\hline
 & 1 &     2322.82x1746\,(7)   &     \\
 & 2 &     2322.82x8684\,(10)  &     \\
 & 3 &     2322.84x038\,(2)    &       \\
 & 4 &     2322.85x698\,(3)    &     19x.\,(18.) \\
 & 5 &     2322.87x866\,(5)    &     20x.7\,(8.4) \\
 & 6 &     2322.90x560\,(6)    &     21x.5\,(4.6) \\
 & 7 &     2322.93x798\,(9)    &     21x.8\,(2.9) \\
 & 8 &     2322.97x600\,(12)   &     22x.0\,(2.0) \\
 & 9 &     2323.01x98\,(2)     &     22x.0\,(1.4) \\
 &10 &     2323.06x97\,(2)     &     22x.0\,(1.0) \\
 &11 &     2323.12x58\,(2)     &     21x.89\,(77) \\
 &12 &     2323.18x82\,(3)     &     21x.81\,(59) \\
\end{tabular}
\end{ruledtabular}
\end{table}

%
%
\begin{table}
\caption{SE correction to the $g$ factor of $2P_J$ states of hydrogen-like
ions, in ppm.  $I_{nP_J}$ is the higher-order remainder.
 \label{tab:gfact:p} }
\begin{ruledtabular}
\begin{tabular}{cr..}
$2P_{1/2}$ & $Z$ 
& \multicolumn{1}{c}{$\Delta g_{2P_{1/2}}$} 
& \multicolumn{1}{c}{\hskip0.4cm $I_{2P_{1/2}}(\Za)$}  \\
\hline\\[-9pt]
 & 1 &     -774.25x8151\,(3)  &     0.121x258\,(21) \\
 & 2 &     -774.21x2929\,(11) &     0.121x715\,(22) \\
 & 3 &     -774.13x687\,(2)   &     0.122x414\,(19) \\
 & 4 &     -774.02x917\,(3)   &     0.123x280\,(14) \\
 & 5 &     -773.88x876\,(3)   &     0.124x305\,(10) \\
 & 6 &     -773.71x442\,(3)   &     0.125x473\,(7) \\
 & 7 &     -773.50x460\,(3)   &     0.126x803\,(5) \\
 & 8 &     -773.25x838\,(3)   &     0.128x186\,(4) \\
 & 9 &     -772.97x356\,(3)   &     0.129x711\,(2) \\
 &10 &     -772.64x862\,(3)   &     0.131x336\,(2) \\
 &11 &     -772.28x175\,(4)   &     0.133x054\,(3) \\
 &12 &     -771.87x108\,(8)   &     0.134x858\,(5) \\
\hline
\hline\\[-9pt]
$2P_{3/2}$ & $Z$ 
& \multicolumn{1}{c}{$\Delta g_{2P_{3/2}}$} 
& \multicolumn{1}{c}{\hskip0.4cm $I_{2P_{3/2}}(\Za)$}  \\
\hline\\[-9pt]
 & 1 &      774.29x1470\,( 3) &     0.148x104\,(21) \\
 & 2 &      774.34x6522\,(12) &     0.148x294\,(24) \\
 & 3 &      774.43x854\,(3) &     0.148x567\,(24) \\
 & 4 &      774.56x774\,(4) &     0.148x851\,(22) \\
 & 5 &      774.73x495\,(6) &     0.149x338\,(18) \\
 & 6 &      774.94x028\,(6) &     0.149x816\,(14) \\
 & 7 &      775.18x442\,(6) &     0.150x350\,(11) \\
 & 8 &      775.46x799\,(6) &     0.150x933\,(7) \\
 & 9 &      775.79x167\,(5) &     0.151x561\,(4) \\
 &10 &      776.15x615\,(5) &     0.152x231\,(4) \\
 &11 &      776.56x218\,(6) &     0.152x940\,(4) \\
 &12 &      777.01x055\,(6) &     0.153x685\,(4) \\
\end{tabular}
\end{ruledtabular}
\end{table}

%
%
\begin{figure}[thb]
\includegraphics[width=0.9\linewidth]{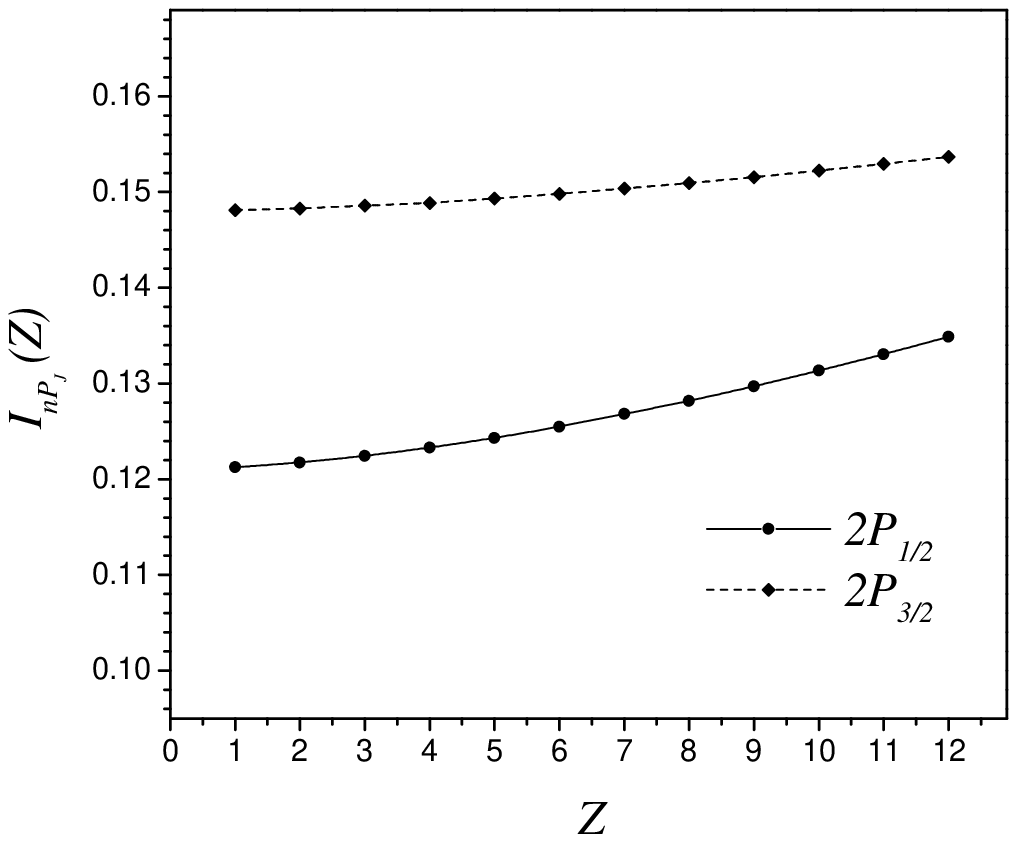}
\caption{The higher-order remainder for the SE correction to the $g$ factor
of the $2P_{1/2}$ and  $2P_{3/2}$   states. 
\label{fig:hop:gfact}}
\end{figure}

%
%
\section{Numerical results}
\label{numerical}

Our calculation of the SE correction to the hfs and the $g$ factor of
hydrogen-like ions was performed in the Feynman gauge and for a
point nucleus. The fine structure constant of $\alpha^{-1} = 137.036$ 
was used in the calculation. 
The small deviation of this value from the currently accepted 
one~\cite{mohr:08:rmp} does not influence the numerical
results for the higher-order remainder. 
An example set of individual contributions to the SE correction to the $1S$ hfs of
atomic hydrogen is presented in Table~\ref{tab:breakdown}. 

The SE correction to the hfs of an $nS$ state can be represented as
\begin{align} 
\label{ee2}
& \Delta E_{nS} = 
\frac{\alpha}{\pi}\, E_F(nS)
  \Biggl[a_{00}+ (\Za)\,a_{10} \nonumber \\ 
& + (\Za)^2\biggl\{ \ln^2[(Z\alpha)^{-2}] \, a_{22}+ 
\ln[(Z\alpha)^{-2}] \,a_{21} +a_{20} \biggr\}
\nonumber \\ 
& + (\Za)^3\,\ln[(Z\alpha)^{-2}] \,a_{31}+ (\Za)^3\,F_{nS}(\Za)
  \Biggr]\,,
\end{align}
where $E_F(nS)$ is the non-relativistic hfs value, and the
$a_{ij}$ are coefficients of the $Z\alpha$-expansion with the 
first index corresponding to the power of $Z\alpha$ and the 
second corresponding to the power of the logarithm.
We have
\begin{align}
a_{00}(nS) =& \; 1/2 \,,  \quad a_{10}(nS) =  -8.03259003 \,,
\nonumber\\[2ex]
a_{22}(nS) =& \; -2/3\,, \quad a_{31}(nS) = -13.30741592\,,
\end{align}
\begin{align}
a_{21}(1S) =& \; -1.334503593\,,
\nonumber\\[2ex]
a_{21}(2S) =& \;  0.317103926 \,,
\nonumber\\[2ex]
a_{21}(3S) =& \;  0.921048823 \,,
\end{align}
\begin{align}
a_{20}(1S) =& \; 17.12233875\,,
\nonumber\\[2ex]
a_{20}(2S) =& \; 11.90110542\,,
\nonumber\\[2ex]
a_{20}(3S) =& \; 10.41704775\,,
\end{align}
see the recent articles
\cite{pachucki:96:pra,nio:97,karshenboim:02:epjd} and references therein for earlier
studies. $F_{nS}$ is the higher-order remainder, which we 
address in our numerical all-order approach. 
Our numerical results for the SE correction to
the hfs of the $1S$, $2S$, and $3S$ states are listed in Table~\ref{tab:hfs:s}.

The SE correction to the hfs of an $nP_J$ states is much less studied. Only
the leading term of its $\Za$ expansion is known today. The correction,
therefore, is written as
\begin{align} \label{ee3}
  \Delta E_{nP_J} &\ = E_F(nP_J)\, \frac{\alpha}{\pi}\,
    \Bigl[a_{00}+ (\Za)^2\,G_{nP_J}(\Za)
         \Bigr]\,,
\end{align}
with $G_{nP_J}$ being the higher-order remainder. The coefficient $a_{00}$ is
given by $a_{00}(nP_{1/2}) = 1/4$ and $a_{00}(nP_{3/2}) = -1/8$ \cite{BrPa1968}. 
Our numerical results for the SE correction to
the hfs of the $2P_{1/2}$ and $2P_{3/2}$ states are listed in Table~\ref{tab:hfs:p}.

The results for the higher-order remainders $F_{nS}$ and $G_{nP_J}$ inferred
from our numerical data are plotted in Figs.~\ref{fig:hos} and \ref{fig:hop},
respectively.  For the $2P_{1/2}$ state, a fit of our results is consistent
with the $\Za$ expansion of the form
\begin{align} \label{ee4}
G_{nP_{1/2}}(\Za) = a_{21} \ln[(\Za)^{-2}] + a_{20}+ \ldots \,,
\end{align}
where the value of the logarithmic coefficient is very close to
$a_{21}(2P_{1/2})=-3/2$ and the constant term is about
$a_{20}(2P_{1/2})=3.5$. For the $2P_{3/2}$ state, the 
numerical data are consistent with $a_{21}(2P_{3/2})= 0$.
Our results in Table~\ref{tab:hfs:p} 
are in moderate agreement with those obtained previously
in Refs.~\cite{sapirstein:06:hfs,sapirstein:08:hfs}
but significantly improve upon them in numerical accuracy. 
Nevertheless, we disagree with the suggestion \cite{sapirstein:06:hfs}
about the possible presence of the squared
logarithm in the $\Za$ expansion (\ref{ee4}) for the $2P_{1/2}$ state. 
A more careful investigation of the analytic structure 
of the higher-order terms is performed in the follow-up paper~\cite{follow-up}.

The SE correction to the bound-electron $g$ factor of an $nS$ state can be
represented as
\begin{align} \label{ee30}
  \Delta g_{nS} =& \; \frac{\alpha}{\pi}\,
    \biggl[1+  \frac{(\Za)^2}{n^2}\,b_{20} 
\nonumber \\ 
& \; + \frac{(\Za)^4}{n^3}
\biggl\{ \ln[(Z\alpha)^{-2}]\,b_{41} + b_{40}\biggr\}
\nonumber \\ & + \frac{(\Za)^5}{n^3}\,H_{nS}(\Za) \biggr]\,,
\end{align}
where the $b_{ij}$ are known coefficients of the $\Za$ expansion:
\begin{align}
b_{20}(nS) =& \; \frac16\,, \qquad b_{41}(nS) = \frac{32}{9}\,, 
\nonumber\\[2ex]
b_{40}(1S) =& \; -10.23652432 \,, 
\nonumber\\[2ex]
b_{40}(2S) =& \; -10.70771560 \,,
\nonumber\\[2ex]
b_{40}(3S) =& \; -11.52963397 \,,
\end{align}
see Ref.~\cite{pachucki:04:prl} and references
therein. $H_{nS}$ is the remainder incorporating all higher-order contributions. It
is remarkable that  the higher-order remainder $H_{nS}$ enters in the
relative order $(\Za)^5$ rather than in the relative order $(\Za)^3$, as in the
case of the hfs. This means that cancellations in extracting the remainder from
numerical results for $Z=1$ are larger for the $g$ factor than for the hfs
by four orders of magnitude.

Our numerical results for the SE correction to the $g$ factor of the electron
$1S$, $2S$, and $3S$ states of light hydrogen-like ions are presented in
Table~\ref{tab:gfact:s}. We observe that the higher-order remainder behaves
very similarly for all $nS$ states studied, the $2S$ remainder being just
about 2\% larger than that for the $1S$ states and the $3S$ and $2S$
remainders being equal within the numerical
uncertainty. The accuracy of the direct
numerical determination of the $1S$ remainder for $Z=1$ and $Z=2$ can easily
be increased by extrapolating values obtained for higher values of $Z$. An
extrapolation yields the improved results $H_{1S}(1\alpha) = 23.39\,(80)$
and $H_{1S}(2\alpha) = 23.03\,(44)$. Improved values of the
higher-order remainder for the $2S$ and $3S$ states are most easily obtained
by scaling the $1S$ remainder.
The trend of the higher-order remainder for low $Z$ 
is consistent with a numerically large, $n$-independent coefficient 
$b_{50}$ in Eq.~(\ref{ee30}).

For the $P_J$ states, the bound-electron $g$ factor is
studied less thoroughly than for the $S$ states. The leading term of its
$\Za$ expansion is due to the electron anomalous magnetic moment (amm) and is
immediately obtained for a general state as \cite{grotch:73:kashuba}
\begin{align} \label{ee30a}
  b_{00} = \frac{1-2\kappa}{4j(j+1)}\,,
\end{align}
where $\kappa$ is the Dirac quantum number and $j$ is the total angular
momentum of the electron state. For the $P$ states, the
explicit results are $b_{00}(nP_{1/2}) = -1/3$ and 
$b_{00}(nP_{3/2}) = 1/3$. 

The next-order term, $b_{20}$, consists of two
parts, one induced by the electron amm and
the other, by the emission and the absorption of virtual photons of low energy
(commensurate with the electron binding energy). A simple calculation of the
first part gives \cite{grotch:73:kashuba} $b_{20}(nP_{1/2},{\rm amm}) = -1/2$
and $b_{20}(nP_{3/2},{\rm amm}) = 1/10$. The second part is
nonvanishing for states with $l\ne 0$ only and is more complicated. Its
general expression is known \cite{grotch:73:hegstrom,pachucki:04:lwqed}
but the only numerical result available for hydrogenic atoms is the
estimate made in Ref.~\cite{calmet:78}, which disagrees 
with our numerical values both in the sign and the magnitude. 
Commenting on this fact, we note that
the estimate is based on a rather crude
approximation. Namely, the sum over the entire discrete
and continuous spectrum of virtual states was
replaced by the contribution of the lowest 12 discrete bound states only. 
We argue that such approximation might be inapplicable for the problem in hand.
The reason is that, e.g., for the Bethe
logarithm (which is also a contribution induced by the low-energy photons)
the dominant contribution originates from the continuum spectrum
\cite{drake:99:cjp} so that such approximation is clearly inadequate. 

Since the $(\Za)^2$ term is not presently known anyalytically,  
we define the higher-order remainder for the $P_J$ states as
\begin{align} \label{ee31}
  \Delta g_{nP_J}  = \frac{\alpha}{\pi}\,
    \Bigl[b_{00} + (\Za)^2\,I_{nP_J}(\Za)
         \Bigr]\,.
\end{align}
Our numerical results for the SE correction to the $g$ factor of the electron
$2P_{1/2}$ and $2P_{3/2}$ states of light hydrogen-like ions are listed in
Table~\ref{tab:gfact:p}. The corresponding higher-order remainder function is
plotted in Fig.~\ref{fig:hop:gfact}. 

%
%
\section{CONCLUSIONS}
\label{conclusions}

We have discussed, in detail, a numerical evaluation,
nonperturbative in the binding Coulomb field, of the self-energy 
correction to the hyperfine splitting and of the 
self-energy correction to the $g$ factor in 
hydrogen-like ions with low nuclear charge number $Z=1,\dots,12$. 
We consider the ground state, as well as the $2S$ and $3S$ excited states,
and the $2P_{1/2}$ as well as $2P_{3/2}$ states.
The value of $\alpha^{-1} =137.036$ is employed in all
calculations. At the level of precision we are operating at, 
the final results for the self-energy corrections
depend on the precise value employed very sensitively.
However, the main dependence on the value of $\alpha$ is
accounted for by the analytically known lower-order 
terms. Thus, the results for the remainder functions 
$F_{nS}(\Za)$, $G_{nP_J}(\Za)$,
$H_{nS}(\Za)$, and $I_{nP_J}(\Za)$ as given in 
Tables~\ref{tab:hfs:s},~\ref{tab:hfs:p},~\ref{tab:gfact:s} 
and~\ref{tab:gfact:p} are not influenced by the value of 
$\alpha$ employed. Even if a value of $\alpha$ which 
differs from $\alpha^{-1} =137.036$ on the level 
$10^{-7}$ were employed, then the values of the 
remainder functions would not change: their main 
uncertainty is due to limits of convergence of the 
integrals that constitute the 
nonperturbative self-energy corrections,
as described in the preceding sections of this 
article.

The organization of our calculation is described 
in Sec.~\ref{detailed}. We consider separately the 
reference-state contribution to the reducible and the 
vertex part (Sec.~\ref{refstatecontrib}) and
the zero-potential contribution to the 
vertex and to the reducible part (Sec.~\ref{zeropotredver}).
The one-potential and many-potential vertex parts,
which represent the most challenging part of the calculation,
are discussed in
Secs.~\ref{onepotver} and~\ref{manypotver}.
The many-potential reducible part and the 
irreducible part conclude the discussion of our 
computational method 
(Secs.~\ref{manypotred} and~\ref{irreducible}).
Numerical calculations were carried out on the parallel
computing environments of MPI Heidelberg and MST Rolla.

It is instructive to compare the numerical 
results obtained (Tables~\ref{tab:hfs:s}---\ref{tab:gfact:p})
to analytic results from the $\Za$ expansion.
The analytic parameterization of the self-energy correction to the 
hyperfine splitting according to Eq.~(\ref{ee2})
entails both logarithmic as well as nonlogarithmic 
corrections. Our numerical results for the scaled
self-energy correction $\delta E_n$ to the hyperfine
splitting and for the nonperturbative remainder function 
$F_{nS}(\Za)$ are given in Table~\ref{tab:hfs:s} for $S$ states.
A plot of the data (see Fig.~\ref{fig:hos}) indicates that the 
higher-order remainders $F_{nS}(\Za)$, for 
$Z\to 0$, may converge toward an $a_{30}$ coefficient 
which is significantly dependent on the principal quantum number.

The scaled self-energy correction $\delta E_{nP_J}$ for
$2P_{1/2}$ and $2P_{3/2}$ states is analyzed in Table~\ref{tab:hfs:p}.
A plot of the data (see Fig.~\ref{fig:hop}) aids in a
comparison to an 
analytic model for the correction, given in Eq.~(\ref{ee4}).
The nonperturbative remainder $G_{nP_J}(Z\alpha)$ for the hfs of $P$ states
is seen to be well represented by an analytic  model of the form
$\{ a_{21} \ln[(\Za)^{-2}] + a_{20}+\ldots \}$,
where a fit to the numerical data indicates that
$a_{21}(2P_{1/2})=-3/2$ and that
$a_{21}(2P_{3/2})=0$. It has been suggested in 
Ref.~\cite{sapirstein:06:hfs} that a ``double'' (squared)
logarithm in the $\Za$ expansion could be present
for low nuclear charge; the latter would correspond
to a nonvanishing $a_{22}$ coefficient for $P$ states.
Our numerical results in Table~\ref{tab:hfs:p} 
do not contradict those of Ref.~\cite{sapirstein:06:hfs} on the 
level of numerical accuracy obtained in the cited reference.
However, we cannot confirm the presence of such a 
double-logarithmic correction (see also~\cite{follow-up}).

A large number of analytic terms are known for the 
self-energy correction to the $g$ factor for $S$ states
[see Eq.~(\ref{ee30})]. The 
higher-order remainder $H_{nS}(\Za)$ for the $g$ factor of $S$ states
is thus ``separated'' from the leading-order effect by 
about ten orders of magnitude for $Z=1$.
Thus, although our direct numerical
evaluation of the self-energy correction 
for the ground state at $Z=1$ is precise
[$\Delta g_{1S} = 2322.840230\,(2)  \times 10^{-6}$ 
at $Z=1$], we can only infer the higher-order
remainder $H_{1S}(\Za)$ at $Z=1$ to about $\pm$10\%: 
the result after the subtraction of lower-order terms
is $H_{1S}(1\alpha) = 19\,(3)$. By extrapolation 
of more accurate data for the remainder obtained from higher 
values of the nuclear charge, we can obtain
the improved results $H_{1S}(1\alpha) = 23.39\,(80)$
and $H_{1S}(2\alpha) = 23.03\,(44)$. 
The remainder functions at very low $Z$ appear to depend
only very slightly on the principal quantum number;
they are consistent with $H_{nS}(Z\alpha)$ approaching 
an $n$-independent coefficient $b_{50}$ as $Z \to 0$.

For the $g$ factor of $P$ states, only the leading 
coefficient $b_{00}$ is known from the $\Za$ expansion
[see Eq.~(\ref{ee31})].
The self-energy remainder function
$I_{nP_J}(\Za)$ for the $g$ factor of $P$ states
can be inferred from our numerical data in 
Table~\ref{tab:gfact:p} after subtraction of the 
leading analytic term as given in Eq.~(\ref{ee31}).
The numerical data for $P$ states are consistent with the functions
$I_{nP_J}(\Za)$ tending toward a constant for 
$Z\to 0$. A plot of the data in Fig.~\ref{fig:hop:gfact}
confirms this trend.

To conclude, we have performed an all-order (in $Z\alpha$) calculation of the 
self-energy correction to hyperfine splitting and $g$ factor 
in hydrogen-like ions with low nuclear charge numbers.
The calculation is accurate enough to infer higher-order remainder 
terms without any additional extrapolation, by a simple subtraction
of the known terms in the $Z\alpha$-expansion.
We improve the numerical accuracy by several orders of magnitude as
compared to previous evaluations; this leads to improved 
theoretical predictions for all QED 
effects considered in this article. 

%
%
\section*{Acknowledgments}

Enlightening discussions with K.~Pachucki on the $g$ factor of $P$ states are 
gratefully acknowledged.
V.A.Y.~was supported by DFG (grant No.~436 RUS 113/853/0-1), by
RFBR (grant No.~06-02-04007), and by the foundation ``Dynasty.''
U.D.J.~has been supported by the National Science Foundation (Grant PHY-–8555454)
as well as by a Precision Measurement Grant from the National Institute of
Standards and Technology.



\appendix
\section{Angular integrations in momentum space}
\label{app:1}

In this section we demonstrate how to perform the integration over the angular
variables in momentum space. The problem in hand can be formulated as follows:
the general expression of the form
\begin{equation}
\int d\hp_1\, d\hp_2\,F(p_{1r},p_{2r},\xi)\,A(\hp_1)\,B(\hp_2)\,,
\end{equation}
where $F$, $A$, and $B$ are some arbitrary functions and $\xi = \hp_1\cdot\hp_2$,
needs to be integrated over all angular variables exept for $\xi$, i.e., to be
reduced to the form
\begin{equation}
\int_{-1}^1 d\xi\,F(p_{1r},p_{2r},\xi)\,X(A,B;\xi)\,.
\end{equation}
In order to write a general expression for the function $X(\xi)$ in terms of
$A$ and $B$, we use the standard decomposition of the function $F$ in terms of
the spherical harmonics $Y_{lm}$,
\begin{align}
F(p_{1r},p_{2r},\xi) &\ = 2\pi \sum_{lm} Y^*_{lm}(\hp_1)\,Y_{lm}(\hp_2)
\nonumber \\ & \times
  \,\int_{-1}^1d\xi\, F(p_{1r},p_{2r},\xi)\,P_l(\xi)\,,
\end{align}
where $P_l$ are the Legendre polynomials. From this, we immedately have
\begin{align} \label{app3}
X(A,B;\xi) &\ = 2\pi\, 
\sum_{l=0}^\infty P_l(\xi)\,
\sum_{m=-l}^l
  \left[\int d\hp_1\,A(\hp_1)Y^*_{lm}(\hp_1)\right]
\nonumber \\ & \times
  \left[\int d\hp_2\,B(\hp_2)Y_{lm}(\hp_2)\right]\,.
\end{align}
This general formula greatly simplifies when one of
the functions (say, $B$) is unity or the identity
function, i.e., $B(\hp_2) = \hp_2$.
When $B$ is unity, only the term with $l=0$ 
contributes, and we have:
\begin{equation}
X(A,1;\xi) = 2\pi \int d\hp\,A(\hp)\,.
\end{equation}
When $B = {\rm id}$,
\begin{equation}
X(A, {\rm id};\xi) = 2\pi \xi \int d\hp\,\hp\,A(\hp)\,.
\end{equation}
In more complicated cases with $B = \hp_i\hp_k\ldots$, formulas for $X$ can be
in priciple obtained by using the Racah algebra. Alternatively, one can observe [from
Eq.~(\ref{app3})] that $X$ is a combination of the 
Legendre polynomials with some coefficients. It is straightforward to
find the coefficients by performing integrations in Eq.~(\ref{app3})
analytically for each particular case (where advantage may be taken of
computer algebra).

In the present work, we need angular integrals of three types, $K_1$,
$K_2$, and $K_3$, defined as ($\mu = 1/2$):
\begin{subequations}
\begin{align}
\frac{3i}{4\pi} \int d\hp_1\, d\hp_2\, & \ F(p_{1r},p_{2r},\xi)\,
     \chi^{\dag}_{\kappa, \mu}(\hp_1)\, [\hp_1\times \bsigma]_z\,
      \chi_{-\kappa,\mu}(\hp_2) 
\nonumber \\ &
= \int_{-1}^1 d\xi\,F(p_{1r},p_{2r},\xi)\,
        K_1(\kappa)\,,
\\
\frac{3i}{4\pi} \int d\hp_1\, d\hp_2\, & \ F(p_{1r},p_{2r},\xi)\,
     \chi^{\dag}_{\kappa, \mu}(\hp_1)\, [\hp_2\times \bsigma]_z\,
      \chi_{-\kappa,\mu}(\hp_2) 
\nonumber \\ &
= \int_{-1}^1 d\xi\,F(p_{1r},p_{2r},\xi)\,
        K_1^{\prime}(\kappa)\,,
\\
\frac{3i}{4\pi} \int d\hp_1\, d\hp_2\, & \ F(p_{1r},p_{2r},\xi)\,
     \chi^{\dag}_{\kappa, \mu}(\hp_1)\, [\hp_1\times \hp_2]_z\,
      \chi_{\kappa,\mu}(\hp_2) 
\nonumber \\ &
= \int_{-1}^1 d\xi\,F(p_{1r},p_{2r},\xi)\,
        K_2(\kappa)\,,
\\
\frac{3i}{4\pi} \int d\hp_1\, d\hp_2\, & \ F(p_{1r},p_{2r},\xi)\,
     \chi^{\dag}_{\kappa, \mu}(\hp_1)\, i \bsigma_z\,
      \chi_{\kappa,\mu}(\hp_2) 
\nonumber \\ &
= \int_{-1}^1 d\xi\,F(p_{1r},p_{2r},\xi)\,
        K_3(\kappa)\,.
\end{align}
\end{subequations}
Using the technique described above, we obtain the following results for the
basic angular integrals:
\begin{subequations}
\begin{align}
K_1(\kappa) =& \; \left\{
        \begin{array}{cl}
        -\xi\,, & \kappa = -1\,, \\[2ex]
          1\,, & \kappa = 1\,, \\[2ex]
\displaystyle  \tfrac15 \, (1-3 \,\xi^2)\,, & \kappa = -2\,,\\[2ex]
\displaystyle  \tfrac25 \,  \xi\,, & \kappa = 2\,,
        \end{array}
   \right. \\[2ex]
K_1^{\prime}(\kappa) =& \; \left\{
        \begin{array}{cl}
        -1\,, & \kappa = -1\,, \\[2ex]
          \xi\,, & \kappa = 1\,, \\[2ex]
\displaystyle   -\tfrac25 \xi\,, & \kappa = -2\,,\\[2ex]
\displaystyle   -\tfrac15 (1-3\xi^2)\,, & \kappa = 2\,,
        \end{array}
   \right.\\[2ex]
\nonumber
\end{align}
as well as
\begin{align}
K_2(\kappa) =& \; \left\{
        \begin{array}{cl}
        0\,, & \kappa = -1\,, \\[2ex]
        -\tfrac12 (1-\xi^2)\,, & \kappa = 1\,, \\[2ex]
        -\tfrac14 (1-\xi^2)\,, & \kappa = -2\,,\\[2ex]
     -\tfrac{9}{20} \xi(1-\xi^2)\,, & \kappa = 2\,,
        \end{array}
   \right. \\[2ex]
K_3(\kappa) =& \; \left\{
        \begin{array}{cl}
        -\tfrac32\,, & \kappa = -1\,, \\[2ex]
        \tfrac12 \xi\,, & \kappa = 1\,, \\[2ex]
        -\tfrac12 \xi\,, & \kappa = -2\,,\\[2ex]
     -\tfrac{3}{20} (1-3\xi^2)\,, & \kappa = 2\,.
        \end{array}
   \right.
\end{align}
\end{subequations}

\newpage


\end{document}